\documentclass{aa}  

\usepackage{graphicx}
\usepackage{txfonts}
\usepackage[colorlinks,citecolor=blue,hypertexnames=true,bookmarks=false,breaklinks=true]{hyperref}
\usepackage[dvipsnames]{xcolor}
\usepackage{makecell}
\newcommand{\acosmos}[0]{A$^3$COSMOS}
\newcommand{\agoodss}[0]{A$^3$GOODSS}
\usepackage{placeins}

\begin{document}

   \title{\acosmos{}: The dust content of massive quiescent galaxies and its evolution with cosmic time}
\titlerunning{\acosmos{}: The dust content of massive QGs and its evolution with cosmic time}

   \author{
   Sylvia Adscheid \inst{1}
   \and
   Benjamin Magnelli \inst{2}
   \and
   Laure Ciesla \inst{3}
   \and   
   Daizhong Liu \inst{4}
   \and
   Eva Schinnerer \inst{5}
   \and
   Frank Bertoldi \inst{1}
   }

   \institute{
            Argelander-Institut für Astronomie, Universität Bonn, Auf dem Hügel 71, 53121 Bonn, Germany\\
            \email{sadscheid@astro.uni-bonn.de}
        \and
            Université Paris-Saclay, Université Paris Cité, CEA, CNRS, AIM, 91191, Gif-sur-Yvette, France
        \and
            Aix Marseille Univ, CNRS, CNES, LAM, Marseille, France
        \and
            Purple Mountain Observatory, Chinese Academy of Sciences, 10 Yuanhua Road, Nanjing 210023, China
        \and
            Max Planck Institute for Astronomy, Königstuhl 17, D-69117, Germany
   }

   \date{}

  \abstract
   {}
   {We study the dust content of massive ($\log(M_*/M_{\odot})\geq10.8$) quiescent galaxies (QGs) at redshifts $z=0.5-3$ to place constraints on the evolution of their cold interstellar medium (ISM) and thereby obtain insights into the processes of galaxy quenching throughout cosmic time.}
   {We used a robust sample of 458 colour-selected QGs covered by the \acosmos{}+\agoodss{} database to perform a stacking analysis in the $uv$ domain and measured their mean dust masses from their stacked sub-millimetre luminosities. 
   We used the \texttt{CIGALE} spectral energy distribution fitting code to obtain star formation histories and infer the time since quenching for all the QGs in our sample. 
   We used this information to gain insight into the time evolution of the dust content after quenching.}
   {Most QGs in our sample quenched around a redshift of $z\sim1.3$, following the peak of cosmic star formation.
   The majority of QGs observed at $z>1$ are recently quenched (i.e. quenched for no longer than 500\,Myr), whereas the majority of QGs observed at $z<1$ have already been quenched for a significant amount of time ($\gtrsim1$\,Gyr).
   This implies that high-redshift galaxies ($z\gtrsim2$) are ideal for studying the mechanisms of quenching and its effects on the ISM, while lower-redshift galaxies are more suitable for studying the long-term effects of the QG environment on their ISM. 
   We obtain upper limits on the dust mass fraction of the QG population that indicate a lower dust content in high-redshift massive QGs than what was found by earlier stacking studies, and significantly lower (by a factor of $\sim$2--6) than that of normal star-forming galaxies.
   We also place constraints on the initial gas fraction right after quenching. 
   We find that within the first $\sim600$\,Myr after quenching, QGs already lose on average $\gtrsim70\%$ of their cold ISM.
   Our findings support a gas consumption or removal scenario acting on short timescales.}
   {}


   \maketitle

\section{Introduction}

It has been known for many years that galaxies constitute a distinctly bimodal population that can be observed from the local Universe out to at least cosmic noon \citep[$z\sim2$; e.g.][]{williams09,wuyts11}. This population is divided into 
(i) active star-forming galaxies (SFGs), which are disk-dominated systems with young stellar populations that follow a tight relation in the star formation rate (SFR)--stellar mass plane, the so-called main sequence \citep[MS; e.g.][]{noeske07, speagle14,schreiber15,popesso23,koprowski24}, 
and (ii) passive quiescent galaxies (QGs), which are bulge-dominated systems with old stellar populations and relatively high stellar masses that are located well below the MS due to very low SFRs.
QGs can have high stellar masses of $\log(M_*/M_{\odot})>11$.
To accumulate their stellar mass, they must in the past have been part of the SFG population.
The absence of a significant population between SFGs and QGs \citep[this scantly populated transit region is commonly referred to as the `green valley'; e.g.][]{salim14} indicates that the transition from the star-forming (SF) phase to quiescence (`quenching') must occur on short timescales. 
The observation of QGs at high redshifts of $z\sim3-7$ with the \textit{James Webb} Space Telescope \citep[JWST; e.g.][]{valentino23,carnall23,degraaf2024,weibel24} adds to this conundrum as it poses even more pressingly the question as to how these objects stopped forming stars after assembling high stellar masses within a short cosmic time. 
Understanding which mechanisms drive the quenching of galaxies is therefore central to the understanding and modelling of galaxy evolution.

The formation of stars is tightly linked to the interstellar medium (ISM) of a galaxy because stars form from clouds of cold molecular gas \citep[e.g.][]{shu87}. 
Quenching must therefore efficiently prevent galaxies from forming stars from their cold gas reservoir, by stabilising the gas against gravitational collapse \citep[morphological quenching; e.g.][]{martig09,lin19,lesniewska23,michalowski24}, by dissociating it through heating by active galactic nuclei (AGNs) or stellar feedback \citep[e.g.][]{conroy15,li20,arjona24}, or by removing it from the galaxy.
Possible mechanisms for the removal of cold gas from a galaxy include ejection through outflows driven by AGNs \citep[e.g.][]{page12,cicone14,carniani16,belli24} 
or stellar feedback \citep[e.g.][]{cicone14,hopkins14,dome24}, and consumption through star formation combined with the prevention of fresh gas accretion \citep[starvation; e.g.][]{larson80,feldmann15,boselli16,trussler20}.
However, it remains unclear as to which of these mechanisms lead to or dominate galaxy quenching throughout cosmic time.
To address this question, it is necessary to determine the cold gas content of QGs and the timescales over which it evolves. 
A commonly used tracer of the cold ISM is the cold dust within it that is observable at far-infrared (FIR) and (sub-)millimetre wavelengths; the latter probes the Rayleigh-Jeans spectral regime.

Quiescent galaxies  have very low gas and dust content \citep[e.g.][]{young11,boselli14,michalowski19,michalowski24,lesniewska23} and are therefore difficult to detect. 
Furthermore, converting between dust and molecular gas measurements is complicated by the uncertain gas-to-dust ratio (GDR) in QGs. 
While a GDR of the order of $\sim100$, corresponding to a typical GDR in SFGs at solar metallicity \citep[e.g.][]{magdis12}, is usually assumed, simulations suggest that the GDR in QGs could vary by up to four orders of magnitude \citep[][]{whitaker21b}.
Despite these limitations, two main approaches exist to study the QG dust emission at high redshifts: (i)
deep observations of individual (sometimes lensed) QGs with the Atacama Large Millimeter/submillimeter Array \citep[ALMA; e.g.][]{bezanson19,williams21,whitaker21,gobat22} and (ii) the stacking of larger samples from wide areas surveys, such as those conducted with \textit{Herschel} and the \textit{James Clerk Maxwell} Telescope \citep[JCMT;][]{gobat18,magdis21}.
These two approaches have led to somewhat conflicting results, with stacking analyses finding dust-to-stellar mass fractions of the order of $\sim9\cdot10^{-4}$, and the individual observations pointing towards lower dust mass fractions of $\lesssim4\cdot10^{-4}$. 
\citet{blanquez23} attempted to resolve this discrepancy by stacking a sample of 121 QGs using a selection of observations of the GOODS-South field from the ALMA archive. 
Their results reduce, but do not resolve, the tension between FIR stacking and individual observations, favouring the FIR stacking results.
However, their sample included QGs below the commonly used stellar mass limit of $10^{10.8}\,M_{\odot}$ and did not take the mass dependence of the dust mass fraction  into account.  
The origin of the tension between the different dust mass estimates thus remains unclear. 
To address this, it may help to expand the stacking analysis to a larger and mass-matched sample of massive QGs. 

In this work we present a stacking analysis of a large and mass-complete sample of colour-selected massive QGs using data from the Automated Mining of the ALMA Archive in the COSMOS and GOODS-South Field \citep[\acosmos{}/\agoodss{};][]{a3cosmos_1,adscheid24}, which we used to measure the dust content of high-redshift QGs on a population-wide scale. 
We supplemented this with measurements of the star formation history (SFH) through spectral energy distribution (SED) fitting using the \texttt{CIGALE} code \citep{boquien19cigale}.
This provides constraints on the evolution of the dust content after quenching.
Thereby we can place the as yet most stringent upper limits on the gas and dust content of massive QGs ($M_*>10^{10.8}\,M_{\odot}$), showing that they lose the bulk of their cold ISM ($\gtrsim70\%$) already at the beginning of quiescence ($\lesssim600$\,Myr after quenching). 

We present the galaxy selection and ALMA data in Sect.~\ref{sec_data}.
The SED fitting and stacking methods are described in Sect.~\ref{sec_method}.
In Sect.~\ref{sec_results} we present and discuss our results, and we summarise our findings in Sect.~\ref{sec_summary}. 
In the following we assume a flat $\Lambda$ cold dark matter cosmology with $H_0 = 70$\,km\,s$^{-1}$\,Mpc$^{-1}$, $\Omega_{\Lambda}$ = 0.7, $\Omega_M = 0.3,$ and a \citet{salpeter55} initial mass function, correcting stellar masses and SFRs where necessary.

\section{Data}  \label{sec_data}

\subsection{ALMA data}

For the stacking analysis, we made use of the combined \acosmos{}/\agoodss{} database \citep[data version 20220606;][]{adscheid24}. 
This dataset comprises all observations from the public ALMA archive within the Cosmic Evolution Survey field \citep[COSMOS;][]{cosmos} and Great Observatories Origins Deep Survey
Southern field \citep[GOODS-South;][]{dickinson03} publicly available as of 06.06.2022. 
Calibration of the data was performed using the Common Astronomy Software Application \citep[CASA;][]{casa} using the respective \texttt{scriptforPI.py} provided by the ALMA observatory for each project. 
We used all observations from ALMA bands 3 to 7 (0.8 - 3.6\,mm) conducted in Cycle 3 or later. 
Projects from earlier cycles had to be excluded, as the definition of the visibility weights assigned during calibration differs between cycles $\geq3$ and earlier ones, which renders the combination of the visibilities during stacking inaccurate \citep[see][]{wang22}. 
In total, this yields 196 unique projects with 3257 observations. 
Most of these are in the bands 7 and 6 (1505 and 1055, respectively), followed by bands 3 and 4 (470 and 215), while band 5 contains only a few observations (12). 
For details on the data, see \citet{a3cosmos_1} and \citet{adscheid24}.

\subsection{A robust sample of quiescent galaxies}

\begin{figure*}
\centering
\includegraphics[width=0.9\linewidth]{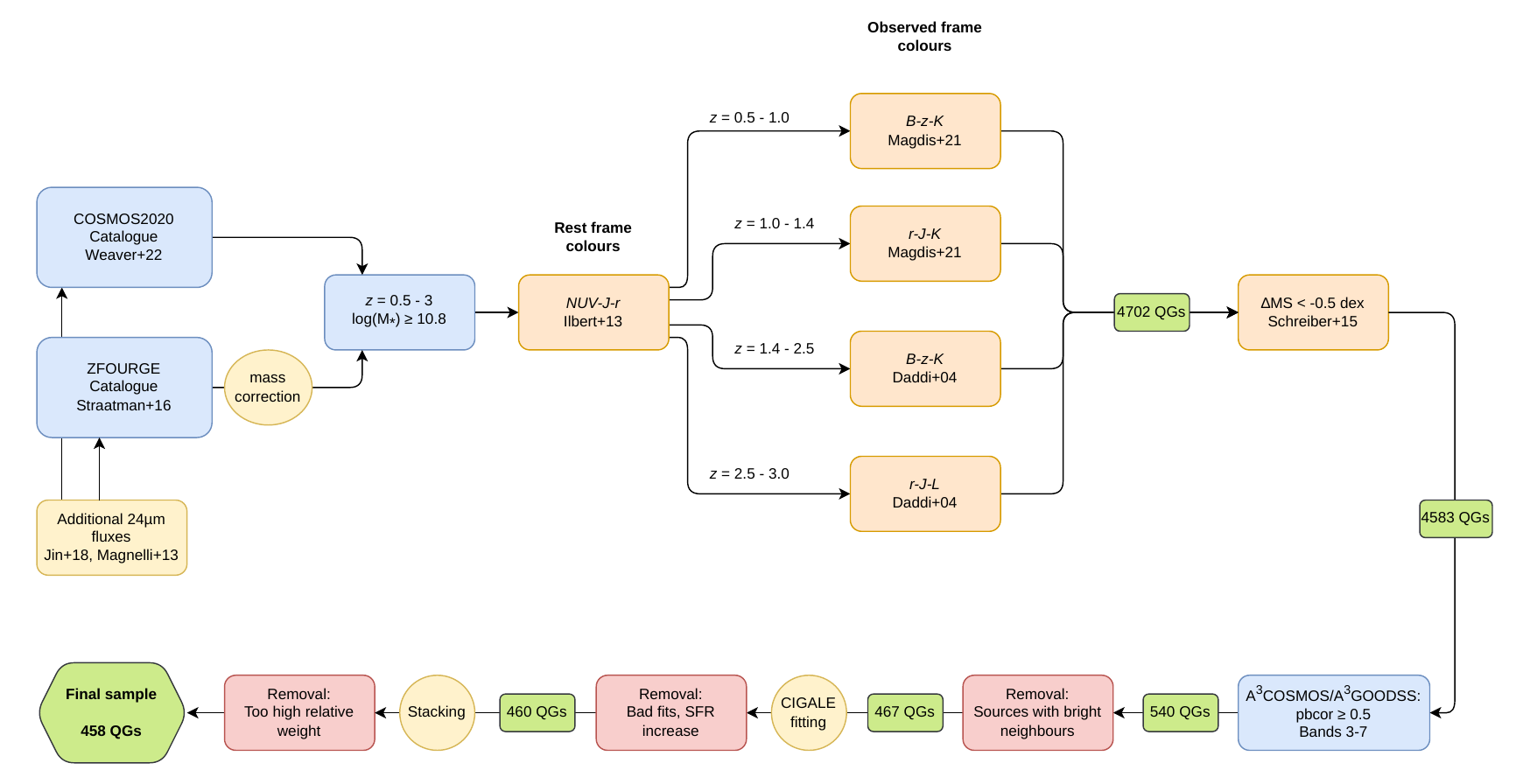}

\caption{Flow diagram of the selection of a robust QG sample. Blue panels denote general selections, orange panels the selection of QGs, red panels further cleaning steps, and green panels  the number of QGs after every step.}
          \label{fig_flowchart}
\end{figure*}

In the following, we describe the process of selecting a robust QG sample within the \acosmos{}/\agoodss{} database.
This process is also schematically outlined in Fig.~\ref{fig_flowchart}.

\subsubsection{Galaxy catalogues}

We selected our sample of QGs from two catalogues: the \textsc{COSMOS2020 Classic} catalogue from \cite{cosmos2020} for the COSMOS field, and the the \textsc{FourStar} galaxy evolution survey (\textsc{ZFOURGE}) catalogue from \cite{straatman16} for GOODS-South. 
Both catalogues provide UV/optical to near-infrared photometry, as well as photometric redshifts and stellar population properties inferred from SED fitting. 
From these catalogues, we selected galaxies (identified by the flags \texttt{lp\_type = 0} in COSMOS2020 and \texttt{Use = 1} in ZFOURGE, respectively) in the redshift range $z=0.5-3$ and with high stellar masses of $\log(M_*/M_{\odot})\geq10.8$.
This particular mass limit was chosen to be comparable with previous works on massive QGs \citep[e.g.][]{gobat18,magdis21}, and was also considered sufficient since, despite the wealth of ALMA observations used, even these massive systems yielded only upper limits in our analysis.
In this high mass regime, both the \textsc{COSMOS2020 Classic} and \textsc{ZFOURGE} catalogue are mass-complete \citep[see][]{straatman16,weaver23}. 
Furthermore, this relatively high mass limit of $10^{10.8}\,M_{\odot}$ also minimises the impact that any potential mass dependence on the dust content of QGs \citep[e.g.][]{blanquez23} could have on our results.

In the creation of the catalogues, different SED-fitting codes were used to infer the redshifts and stellar population properties:
for \textsc{COSMOS2020}, both properties were inferred using the \texttt{LePhare} code \citep{arnouts02lephare,ilbert06lephare}, while for \textsc{ZFOURGE}, the \texttt{EAZY} code \citep{brammer08eazy} was used to obtain the photometric redshifts, and the \texttt{FAST} code \citep{kriek09fast} was used for the stellar population properties. 
To ensure that the samples from the two catalogues are combinable, we used the \textsc{ZFOURGE} catalogue available in the COSMOS region of the Cosmic Assembly Near-infrared Deep Extragalactic Legacy Survey \citep[][]{grogin11} and spatially cross-matched it with the \textsc{COSMOS2020} catalogue (with a matching radius of 1\,arcsec) to identify common sources. 
Restricting this set of common sources to those identified as QGs (38 galaxies; see Sect.~\ref{subsec_colour_selection}), we found very consistent photometric redshifts between these two catalogues (with a median deviation of $<1\%$) but a small systematic offset between their stellar masses, with those in \textsc{ZFOURGE} being 0.11\,dex (median) lower than those in \textsc{COSMOS2020}.
We therefore applied a small upward correction of 0.11\,dex to the stellar masses of all \textsc{ZFOURGE} galaxies to ensure that the masses in the two catalogues are comparable.

In addition to the photometric information from the \textsc{COSMOS2020} and \textsc{ZFOURGE} catalogues, we also included mid-infrared (MIR) photometry from \textit{Spitzer}/Multiband Imaging Photometer for \textit{Spitzer} (MIPS) 24\,$\mu$m to obtain a wider photometric coverage for our SED-fitting (see Sect. \ref{subsec_cigale_fitting}).
These were taken from the `super-deblended' FIR catalogue of \citet{jin18} for COSMOS ($1\sigma\sim16\,\mu$Jy), and from the PEP-GOODS-\textit{Herschel} catalogue of \citet{magnelli13} for GOODS-South ($1\sigma\sim7\,\mu$Jy), which were spatially cross-matched (with a radius of 1\,arcsec) with our galaxy sample. 
Amongst the final 458 galaxies (383 in COSMOS and 75 in GOODS-S) identified as quiescent (see Sects.~\ref{subsec_colour_selection}, \ref{subsec_excluding_neighbours}, \ref{subsec_cigale_fitting}, and \ref{subsec_stacking}), 73 ($\sim$16\%) are detected in the MIPS 24\,$\mu$m bands. 
The properties of these galaxies are specifically discussed in Appendix~\ref{app_analysis_24}.

\subsubsection{Quiescent galaxy selection} \label{subsec_colour_selection}

Quiescent galaxies were selected from the sample of $z = 0.5 - 3$ and log$(M_*/M_{\odot}) \geq 10.8$ galaxies defined in the previous section using a series of different colour selection criteria. 
Following \citet{magdis21}, we first selected galaxies based on the rest-frame \textit{NUV-J-r} criterion of \citet{ilbert13}, 
\begin{eqnarray}
    NUV-r &>& 3\ (r-J) + 1, \\
    NUV-r &>& 3.1. \nonumber
\end{eqnarray}
Then, following again \citet{magdis21}, we applied additional observed-frame colour selections based on the redshift of any given source to further ensure the quiescence of these galaxies. 
For galaxies at $z<1$, we used the \textit{B-z-K} criterion defined for $0.3 \leq z < 1.0$ galaxies by \citet{magdis21}:
\begin{eqnarray}
    z - K &>& 0.3 \cdot (B-z) - 0.5 \nonumber \\
    z - K &\leq& 0.35 \cdot (B-z) +0.15 \\
    z - K &\leq& 0.9 \cdot (B-z) -1.55. \nonumber
\end{eqnarray}
For $1.0 \leq z < 1.4$ galaxies, we used the \textit{r-J-K} criterion from \citet{magdis21},
\begin{eqnarray}
    J - K &>& 0.2 \cdot (r-J) - 0.3 \nonumber \\
    J - K &<& 0.2 \cdot (r-J) +0.275 \\
    r - J &>& 2.0. \nonumber
\end{eqnarray}
For $1.4 \leq z < 2.5$ galaxies, the \textit{B-z-K} criterion from \citet{daddi04},
\begin{eqnarray}
    (z - K) - (B - z) &<& -0.2 \\
    z - K &>& 2.5. \nonumber
\end{eqnarray}
And for $2.5 < z < 4.0$ galaxies, the \textit{r-J-L} criterion from \citet{daddi04}:
\begin{eqnarray}
    (J-L_{3.6}) - 1.4 (r-J) &<& 0 \\
    J-L_{3.6} &>& 2.5.   \nonumber
\end{eqnarray}
For the $L_{3.6}$-band we used the photometry from \textit{Spitzer}/Infrared Array Camera (IRAC) channel 1, which is centred at $3.55\pm0.75\,\mu$m \citep{irac}.

To further ensure the quiescence of the galaxies in our sample, we additionally applied a SFR selection.
Galaxies were only kept if their SFR and stellar mass places them at least 0.5\,dex below the SF MS at their respective redshift, for which we applied the MS definition of \citet{schreiber15}.
Before cross-matching with the area covered by \acosmos{}/\agoodss{} observations, our sample of robust QGs contains 4583 objects.

We note that some previous works also exclude MIR-detected galaxies from their analyses \citep[e.g.][]{magdis21}, while other works include them \citep[e.g.][]{blanquez23}.
We chose to not exclude MIR-detected galaxies a priori to avoid biasing our sample against galaxies containing an AGN.
We instead discuss the influence of such an exclusion on our results in Appendix~\ref{app_analysis_24}.

\subsubsection{\acosmos{}/\agoodss{} selection} \label{subsec_excluding_neighbours}

For the purpose of stacking, we kept only the 540 QGs covered by the \acosmos{}/\agoodss{} database, with a primary beam attenuation $\geq0.5$. 
We then checked for counterparts of those QGs in the \acosmos{}/\agoodss{} catalogues, finding no such counterparts within the high positional accuracy offered by these two catalogues \citep[$\sigma_{\mathrm{pos}}\approx\theta_{\mathrm{FWHM}}/(2\cdot\mathrm{S/N}_{\mathrm{peak}})$;][]{ball75,condon97}. 
This was to be expected from our stringent quiescence selection criteria and the previously reported faint sub-millimetre emission of QGs \citep[e.g.][]{magdis21}. 
In contrast, we found a number of QGs located in the vicinity of \acosmos{}/\agoodss{} detections (supposedly mostly due to projection effects). 
The emission from these nearby bright sources could significantly impact the faint emission of the QGs we wanted to to study. 
To mitigate the impact of nearby bright sources on our results, we decided to exclude from our sample (i) all QGs located closer than 3\,arcsec to a detected source in the \acosmos{}/\agoodss{} catalogues (i.e. $\mathrm{S/N}_{\mathrm{peak}}\geq4.35$ and 5.4 for prior and blindly detected sources, respectively; see \citealt{a3cosmos_1}; 24 QGs are excluded with this criterion), and additionally (ii) QGs located closer to a bright source than a minimum distance $r_{\mathrm{min}}$ (49 QGs are excluded with this criterion). 
This minimum distance was determined from the $\mathrm{S/N}_{\mathrm{peak}}$ of the nearby detected source following\begin{equation}
    r_{\mathrm{min}}(\mathrm{S/N}_{\mathrm{peak}}) = \frac{1}{a} \ln \left( \frac{\mathrm{S/N}_{\mathrm{peak}}}{b} \right).
\end{equation}
The values $a$ and $b$ were empirically calibrated, so that $r_{\mathrm{min}}(\mathrm{S/N}_{\mathrm{peak}}=4.35\ \mathrm{or}\ 5.4)=3$\,arcsec for prior/blindly detected sources, and $r_{\mathrm{min}}(\mathrm{S/N}_{\mathrm{peak}}=50)=10$\,arcsec. 
This two-fold selection criterion allowed us to exclude both the directly affected area in the vicinity around these bright sources (within a $\sim$1-2 FWHM radius) and the wider area still affected by the presence of energetic sidelobes.
Through variation of these parameters, we verified that the exact values of these parameters do not significantly influence the results presented in this paper.

The final sample consists of 467 QGs, with a mean observed redshift of $\langle z_{\mathrm{obs}}\rangle\sim1.1$. 
This increases by a factor of $\sim$4 the number of QGs with ALMA coverage compared to \citet{blanquez23}, who also performed a QG stacking analysis using ALMA data. 
Many of the ALMA archival data used here are also deeper than in \citet{blanquez23}.
The distributions of our \acosmos{}/\agoodss{}-covered sample in stellar mass, redshift, and SFR are very similar to those of the parent sample of 4583 QGs.
We verified that the two samples follow the same underlying distributions using two-sample Kolmogorov-Smirnov (KS) tests ($p$-value $>0.05$).

\section{Methods} \label{sec_method}

\subsection{CIGALE fitting}  \label{subsec_cigale_fitting}

\begin{table*}[]
    \caption[]{Input parameters of the \texttt{CIGALE} module used for SED and SFH fitting of our QG sample.}   
    \tiny
    \label{tab_cigale_params}
$$
    \begin{tabular}{lc}
    \hline
    &\\
    Delayed-$\tau$ SFH   & \citet{ciesla17} \\
    \hline
    age$_{\mathrm{main}}$ [Myr]   &  [1300, 8600] (5-7 values, individual per redshift bin) \\
    $\tau_{\mathrm{main}}$ [Myr]  &  [2000, 17000] (5 values, linearly spaced) \\
    $t_{\mathrm{q}}$ [Myr]        &  10, 30, 50, 70, and [100, 3900] (24 values, log-spaced) \\
    $r_{\mathrm{SFR}}$            &  [0.001, 100] (20 values, log-spaced), and 1 \\    
    \hline
    &\\
    Stellar population & \citet{bruzual03}\\
    \hline
    Metallicity & 0.02 \\
    \hline
    &\\    
    Dust attenuation & \citet{charlot00}  \\
    \hline
    $A_{\mathrm{V,ISM}}$ & [0.2, 2.2] (15 values, linearly spaced)  \\
    $\mu$ & 0.3 \\
    slope$_{\mathrm{ISM}}$ & $-0.7$\\
    slope$_{\mathrm{BC}}$ & $-0.7$\\
    \hline
    &\\
    Dust emission & \citet{dale14} \\
    \hline   
    $\alpha$ & 1.5 ,2.0, 2.5 \\
    \hline
    \end{tabular}
$$
\tablefoot{The parameter age$_{\mathrm{main}}$ denotes the total length of the SFH until the time of observation. 
	The range of this parameter is constrained by the redshift of a given input galaxy and was adjusted individually for each redshift bin} .
    
\end{table*}

To gain insight into the SFHs of the QGs in our sample, we fitted their combined UV-to-MIR photometry information with the SED fitting code \texttt{CIGALE} \citep{boquien19cigale}.
For the SFH, we used the flexible delayed-$\tau$ model from \citet{ciesla17}, tailored to quenched (or bursty) systems:

\begin{eqnarray}
\mathrm{SFR}(t) \propto
\begin{cases}
t \cdot \exp(-t/\tau_{\mathrm{main}}),& t\leq t_\mathrm{flex} \\
r_{\mathrm{SFR}} \cdot \mathrm{SFR}(t_{\mathrm{flex}}),& t> t_\mathrm{flex}.
\end{cases}
\end{eqnarray}

\noindent In this model, the SFR rises and then declines exponentially. 
At a flexible point in time, $t_{\mathrm{flex}}$, it changes suddenly to a constant value until the time of observation, $t_{\mathrm{obs}}$.
The factor $r_{\mathrm{SFR}}$ denotes the ratio of the SFR after and before this point in time, and can be smaller, greater, or equal to 1, thus allowing for quenching, bursty, or no sudden events. 
With that, \texttt{CIGALE} in principle allows both fast and slow quenching:
fast quenchers would correspond to galaxies with $r_{\mathrm{SFR}}<1$ early in their SFH, while slow quenchers would have a largely undisturbed SFH, with the SFR gradually declining with time, slowly moving these systems below the MS.

Following \citet{ciesla21}, we complemented this SFH module with the stellar population model from \citet{bruzual03}, the dust attenuation model from \citet{charlot00}, and the dust emission model from \citet{dale14}.
The input parameters for these models are listed in Table~\ref{tab_cigale_params}.
Finally, for each fitted galaxy, \texttt{CIGALE} provides the SFH\footnote{The SFHs in \texttt{CIGALE} are normalised to produce a stellar mass of 1\,$M_{\odot}$, and the models are then scaled to the observations, which, as \citet{boquien19cigale} show, is equivalent to scaling the SFH to the level of the observations.} and two sets of stellar population parameters: one obtained with the best-fit SED, and one inferred with a Bayesian-like analysis.
In our analysis, we used the Bayesian output parameters, as this avoids degeneracy effects from the discreteness of the input parameters.

The galaxies in our sample have a good photometric coverage with on average $\sim$37 photometric data points.
The SED fits from \texttt{CIGALE} are overall very good, with the reduced chi-square $\chi^2_{\mathrm{red.}}$ following a log-normal distribution with $\mu=-0.1\pm0.2$.
Nevertheless, we identified six galaxies as 3$\sigma$ outliers, that is, galaxies with $\chi^2_{\mathrm{red.}}\gtrsim3$.
These sources have in common that, while they are well fitted in the optical wavelength regime, they are not well fit in one or several \textit{Spitzer}/IRAC MIR bands, hinting at a possible source blending or misassociation.
As a high $\chi^2_{\mathrm{red.}}$ indicates large uncertainties on the output parameters, we excluded these galaxies from our analysis. 
We also identified one source for which \texttt{CIGALE} yields a value of $r_{\mathrm{SFR}}>1$ (whereas the median $r_{\mathrm{SFR}}$ in our sample is 0.005). 
This source also has mostly low signal-to-noise photometry.
As this casts doubt on the quiescent nature of the galaxy, it was also excluded. 
Thus, in total, seven galaxies were excluded, reducing the sample size to 460. 

To ensure the reliability of the galaxy parameters yielded by \texttt{CIGALE}, and especially $t_{\mathrm{q}}$, which we use extensively in Sects.~\ref{subsec_recently_quenched} and \ref{subsec_dustdestruction}, we performed a mock analysis, following \citet[][see also \citealt{boquien19cigale}]{ciesla21}.
It is described in Appendix~\ref{app_cigale_mock}.
This mock analysis reinforces confidence in the galaxy parameters obtained by \texttt{CIGALE}, and in particular the time since quenching, $t_{\mathrm{q}}$, allowing us to incorporate this property into our scientific analysis.

\begin{figure*}
\centering
\includegraphics[width=0.9\linewidth]{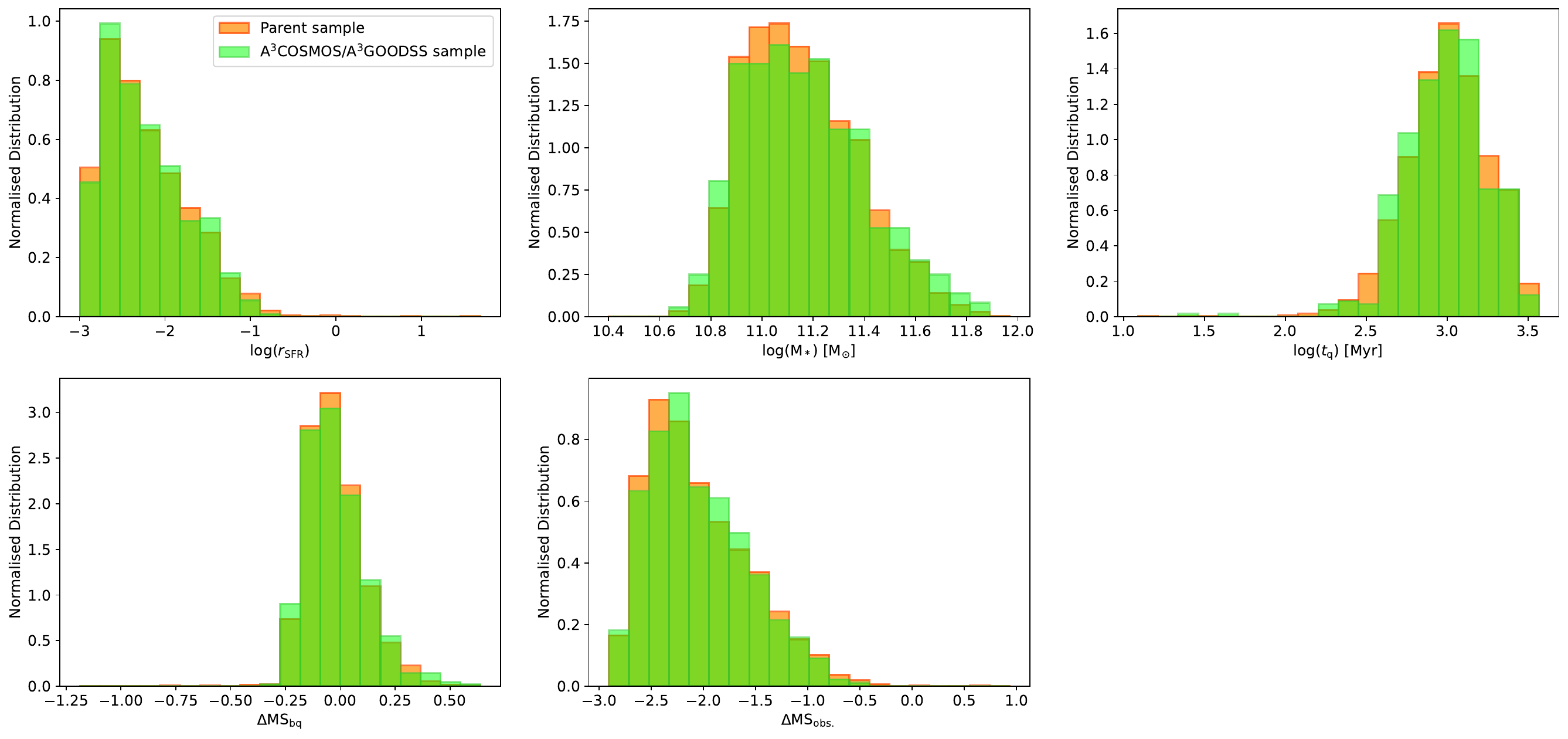}
\caption{Normalised distribution of \texttt{CIGALE}-inferred Bayesian parameters of the parent sample of 4583 QGs and the 460 QGs in our \acosmos{}/\agoodss{} sample: SFR fraction ($r_{\mathrm{SFR}}$), stellar mass ($M_*$), time since quenching ($t_{\mathrm{q}}$), and ($\Delta$MS) before quenching and at the time of observation.
We note that, due to the re-estimation of $M_*$ with \texttt{CIGALE}, a small number of galaxies fall slightly below the initial selection threshold of $\log(M_*/M_{\odot})=10.8$.}
          \label{fig_cigale_histogram}
\end{figure*}

The properties of the 460 QGs in our \acosmos{}/\agoodss{} sample are displayed in Fig.~\ref{fig_cigale_histogram} in comparison to the parent sample of 4583 QGs: $r_{\mathrm{SFR}}$, $M_*$, and $t_{\mathrm{q}}$.
We also show $\Delta$MS, at the time of observation (i.e. $\Delta\mathrm{MS}_{\mathrm{obs}}$) and at the time right before quenching, that is, at $t_{\mathrm{flex}} = t_{\mathrm{obs}} -t_{\mathrm{q}}$:
\begin{equation}
    \Delta\mathrm{MS}_{\mathrm{bq}}= \log\left( \frac{\mathrm{SFR}_{\mathrm{obs}}}{r_{\mathrm{SFR}}\cdot\mathrm{SFR}_{\mathrm{MS}}(z(t_{\mathrm{flex}}),M_*(t_{\mathrm{flex}}))} \right).
\end{equation}
To determine $\mathrm{SFR}_{\mathrm{MS}}(z(t_{\mathrm{flex}}),M_*(t_{\mathrm{flex}}))$ at the cosmic time of the quenching event, we used the redshift-dependent MS relation from \citet{schreiber15}.
The redshift of the quenching event, $z(t_{\mathrm{flex}})$, was inferred using the observed redshift of each source and the time since quenching, $t_{\mathrm{q}}=t_{\mathrm{obs}}-t_{\mathrm{flex}}$, conducting the cosmological calculations with the python package \texttt{astropy.cosmology} \citep{astropy13,astropy18,astropy22}. 
The stellar mass at the time of quenching was computed using $t_{\mathrm{q}}$, $\mathrm{SFR}_{\mathrm{obs}}$, and $M_*(t_{\mathrm{obs}})$, assuming a return fraction $R=0.27$ for a \citet{salpeter55} initial mass function \citep{madau14}:
\begin{equation}
    M_*(t_{\mathrm{flex}}) = M_*(t_{\mathrm{obs}})-(\mathrm{SFR}_{\mathrm{obs}}\cdot t_{\mathrm{q}})\cdot(1-R).
\end{equation}

\noindent Several important conclusions can be drawn from Fig.~\ref{fig_cigale_histogram}.
As per selection the QGs in our sample are massive ($\log(\langle M_*\rangle/M_{\odot})\sim11.2$).
Also, the galaxies in our sample constitute a robust sample of QGs, located well below the MS at the time of observation ($\langle\Delta\mathrm{MS}_{\mathrm{obs}}\rangle_{\mathrm{med}} = -2.14\substack{+0.59\\-0.38}$; interval indicating the 16th and 84th percentiles) and that these systems were, as expected, located around the MS before quenching ($\langle\Delta\mathrm{MS}_{\mathrm{bq}}\rangle_{\mathrm{med}} = 0.04\substack{+0.14\\-0.11}$). 
The majority of our QGs have low $r_{\mathrm{SFR}}$, with 99.6\% of QGs having $r_{\mathrm{SFR}} < 0.1$, and 71\% having $r_{\mathrm{SFR}} < 0.01$, suggesting that our sample is dominated by fast-quenched galaxies.
However, the time since quenching, $t_{\mathrm{q}}$, of many QGs in the sample is long (i.e. $t_{\mathrm{q}}\gtrsim1$\,Gyr) and to conclude that these old systems underwent a rapid quenching event solely on the basis of their low $r_{\mathrm{SFR}}$ values could be a misinterpretation. 
Indeed, while for these systems quenching appears to have begun $t_{\mathrm{q}}$ ago relative to their main smoothly evolving SFH component, the \texttt{CIGALE} module is insensitive to gradual evolutions after this point in time. 
We therefore refrain from making a firm statement about the fraction of fast and slow quenchers in our sample. 
Finally, we note that the relative distributions of these parameters in our QG sample closely mirror those of the parent sample. 
In fact, a two-sample KS test on these distributions indicates that our final sample is consistent with being drawn randomly from our parent sample ($p$-value $>0.05$).
Hence, reducing the QG sample to the areal coverage of \acosmos{}/\agoodss{} thus does not introduce any significant bias in the analysis of the properties of this population.

\subsection{Stacking} \label{subsec_stacking}

To constrain the dust content of our high-redshift, massive QG sample, we performed a stacking analysis in the \textit{uv} domain using all \acosmos{}/\agoodss{} observations available for them. 
To this end, we used CASA and the method described in \citet{wang22} \citep[see also][]{wang24,magnelli24}, to which we refer for the details of the procedure.
In the first step, the observed ALMA visibilities of each galaxy were rescaled to rest-frame luminosities at 850\,$\mu$m, $L_{850}^{\mathrm{rest}}$. 
This requires prior knowledge of the shape of the IR SED.
We therefore used the massive QG SED template of \citet{magdis21}. 
The rescaling factor $\Gamma^{\rm SED}$ is defined as the ratio of the template rest-frame luminosity at 850\,$\mu$m and at the observed wavelength \citep[Eq. 3 of][]{wang22}:
\begin{equation}
    \Gamma^{\rm SED} = L^{\rm SED}_{850} / L^{\rm SED}_{\lambda_{\rm obs}/(1+z)}.
\end{equation}
This factor rescales the ALMA visibility amplitudes, $|V(u,v,w)|_{\lambda_{\rm obs}}$, to 850\,$\mu$m luminosities using Eq. 4 of \citet{wang22}:
\begin{equation}
    |L(u,v,w)|^{\rm rest}_{850} = 4 \pi D_L^2 \cdot |V(u,v,w)|_{\lambda_{\rm obs}} \cdot \Gamma^{\rm SED} / (1+z), 
\end{equation}
where $D_L$ denotes luminosity distance.
The rest-frame wavelengths in our galaxy sample are mostly well in the Rayleigh-Jeans spectral range of the dust continuum (see Appendix~\ref{app_restframe_wavelength}). 
In this regime, the dust SED is described by a power law depending on the dust spectral emissivity index $\beta$, for which \citet{magdis21} adopt a value of $1.8$.
Apart from the chosen $\beta$, the rescaling is therefore template-independent.
In addition, for each pointing, the phase centre of the rescaled visibilities were shifted to the coordinates of the respective galaxy. 
The shifted, rescaled visibilities of all galaxies were then concatenated to create one combined measurement set with all galaxies stacked at the phase centre. 
This combined measurement set was then imaged into the final stack image using the CASA task \texttt{tclean}.
The weight of each individual galaxy in the stacked image depends directly on the number of visibilities that correspond to it and on all subsequent conversion factors that were applied to the visibilities (i.e. rest-frame luminosity rescaling, primary beam correction, etc.).
Therefore, the relative contribution of a given galaxy to the stack can be easily calculated and is higher for galaxies covered by deeper observations (i.e. many visibilities from one or several ALMA projects), and lower for galaxies covered by shallower observations (i.e. fewer visibilities). 
Two galaxies had much higher weights (due to long aggregate integration times), and appeared as $>3\,\sigma$ outliers to the log-normal distribution of galaxy weights in our sample. 
Their high weights implied that the final stacks were essentially an imprint of only these two sources, which was undesirable, as we aimed to retrieve the mean gas content of the entire population of QGs.
We therefore excluded these two galaxies from our analysis, which reduced the total size of our QG sample to 458.
As we demonstrate in Appendix~\ref{app_weight_outliers}, the individual constraints on the dust content of these two galaxies do not contradict our results, but rather reinforce them, particularly for the more massive system at $z\sim1.1$.
Indeed, the dust mass of this system is even below that inferred from stacking for the entire population of QGs.

\subsection{Measuring the dust content} \label{subsec_measure_dust}

We measured the signal in our stacked images (i.e. $L_{850}^{\mathrm{rest}}$) via an aperture photometry approach.
The diameter of the aperture was chosen to be the full width at half maximum of the major axis of the synthesised beam for each stack image, that is, between 0.7 and 1.1\,arcsec.
The luminosity was measured within the respective aperture at the phase centre of the stack and then divided by the intensity of the point spread function within the same aperture.
The uncertainty of the measured luminosity at the image centre was inferred from measurements of luminosities with the same aperture within a radius of $\sim$8\,arcsec from the centre, which is the typical primary beam size in Band 7.

To convert the measured $L_{850}^{\mathrm{rest}}$ to dust masses, we used the QG SED template of \citet{magdis21}, which is normalised to a dust mass of $1.11\cdot10^8\,M_{\odot}$.
We divided our $L_{850}^{\mathrm{rest}}$ by the 850\,$\mu$m luminosity of the template to obtain the weighted mean dust mass of the stack in units of $1.11\cdot10^8\,M_{\odot}$. 
We also used the weights of the galaxies in a stack, as defined in Sect.~\ref{subsec_stacking}, to compute the weighted mean stellar mass and redshift of each stack bin. 
The dust mass inferred using this method naturally depends on the dust temperature assumed by \citet{magdis21}.
QGs generally have distinctly lower dust temperatures than MS galaxies, which typically have dust temperatures of $\sim25-35$\,K in our probed redshift range \citep[e.g.][]{schreiber18}.
If not accounted for, this difference in dust temperature can lead to a significant underestimation of the inferred dust mass \citep[e.g.][]{cochrane22}. 
The QG template of \citet{magdis21} therefore assumes a dust temperature of 21\,K.
We also explore the effect an even lower, yet reasonable dust temperature of 17\,K \citep[][]{cochrane22} would have on our results in Sects.~\ref{subsec_analysis_all} and~\ref{subsec_dustdestruction}. 

\section{Results and discussion} \label{sec_results}

\subsection{Quenching across cosmic time} \label{subsec_recently_quenched}

\begin{figure}
\centering
\includegraphics[width=\linewidth]{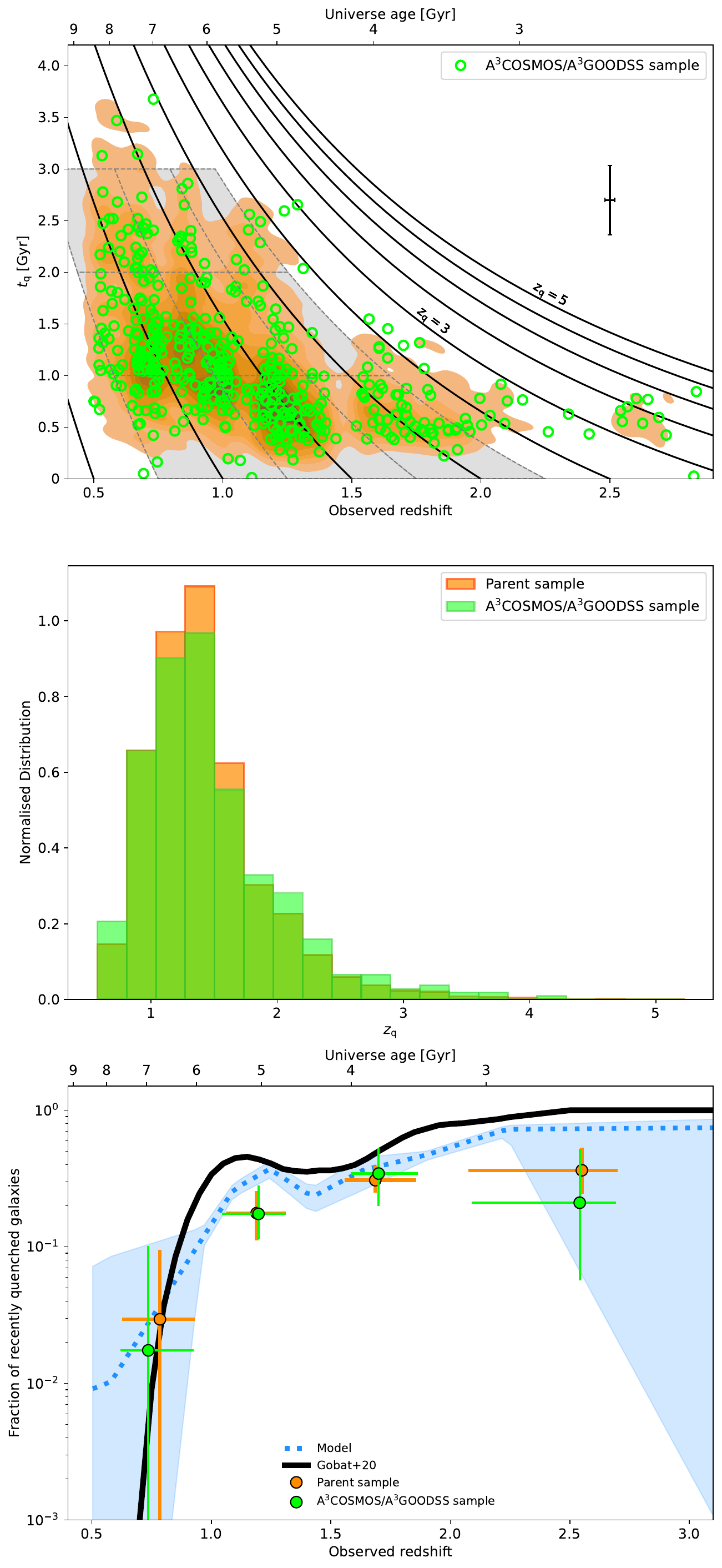}

\caption{Characterisation of our sample of \acosmos{}/\agoodss{} QGs and the parent sample.
\textit{Top panel}: Time since quenching as a function of the observed redshift. 
The median error is shown on the right. 
Solid black lines indicate the locus of galaxies that quenched at the same time, from $z_{\mathrm{q}}=0.5$ to 5 in steps of 0.5. 
The orange density map shows the distribution of the parent sample galaxies. 
The grey shaded areas in the background and dashed grey lines indicate the bins of quenching redshift and time since quenching used in Sect.~\ref{subsec_dustdestruction}.
\textit{Middle panel}: Distribution of quenching redshift ($z_{\mathrm{q}}$).
\textit{Bottom panel}: Fraction of recently quenched galaxies ($t_{\mathrm{q}}\leq500$\,Myr). 
The points and horizontal error bars denote the median redshift and the 16th and 84th percentiles of all galaxies in each bin, respectively. 
The black line shows the fraction of recently quenched massive early-type galaxies as a function of redshift from \citet{gobat20}. 
The dotted blue line and shaded area show our toy model based on the stellar mass functions of \citet{davidzon17}.
}
          \label{fig_combo}
\end{figure}

In the top panel of Fig.~\ref{fig_combo}, we show the time since quenching, $t_{\mathrm{q}}$, as a function of observed redshift for our QGs in comparison to the parent sample.
The distributions of the two samples in this plane are very consistent with one another, with the highest density of galaxies in the range of $z_{\mathrm{obs}}\approx0.5-1.5$ and $t_{\mathrm{q}}\approx0.5-1.5$\,Gyr.
In this diagram, galaxies quenched at the same time (i.e. $z_{\mathrm{q}}$) would fall on a common track, indicated by the solid lines.
From these tracks we can identify galaxies that quenched at a similar time and then evolved for a similar period after the end of their SF activity. 
There are very few galaxies in both samples that quenched at $z_{\mathrm{q}}>3$, and almost none at $z_{\mathrm{q}}>4$ (see also the middle panel of Fig.~\ref{fig_combo}). 
Most QGs in our sample quenched at $z_{\mathrm{q}}\sim0.8-2$, with a maximum around $z_{\mathrm{q}}\sim1.3$.
This is in agreement with earlier studies that found the number and mass density of massive QGs to strongly increase between $z\sim3$ and $z\sim1$ and flatten afterwards \citep[e.g.][]{brammer11,ilbert13,davidzon17}.
This peak in quenching activity closely follows the peak of the cosmic SFR density at $z=1-2$ \citep[e.g.][]{ilbert13}, which suggests a link between these two, with the quenching of these massive systems marking the beginning of the global decline of the cosmic star formation activity. 
Finally, we note that only $\sim$17\% of QGs in our sample are quenched after $z_{\mathrm{q}}<1$ and thus at these redshifts, the majority of massive QGs are old. 

Taken together, all these findings imply that the fraction of recently quenched galaxies within the entire QG population depends on the observed redshift (i.e. $z_{\mathrm{obs}}$), as already pointed out by \citet{gobat20}.
This is clearly demonstrated in the bottom panel of Fig.~\ref{fig_combo}, in which we show the fraction of recently quenched QGs, that is, with $t_{\mathrm{q}}\leq500$\,Myr, as a function of cosmic time, in our \acosmos{}/\agoodss{} and parent sample.
Uncertainties on these fractions were inferred by randomly disturbing the $t_{\mathrm{q}}$ parameter of each QG with its error and recomputing the fraction 1000 times.
The vertical error bars show the 16th and 84th percentiles of these 1000 realisations, with additional Poisson errors added in quadrature, using the Poisson limits from \citet{gehrels86}.
The \acosmos{}/\agoodss{} and parent samples are fully consistent within their errors.
We see a clear increase in the recently quenched fraction by an order of magnitude between $z_{\mathrm{obs}}\sim0.75$ and $z_{\mathrm{obs}}>1$, after which the trend flattens.
Our results are in agreement with predictions of \citet{gobat20} (based on the evolution of the stellar mass function of massive QGs), except at $z_{\mathrm{obs}}>2$, where our recently quenched fraction is significantly lower (i.e. $0.21\substack{+0.31 \\ -0.15}$ compared to $>0.8$). 
This high-redshift discrepancy could stem from our very conservative selection of QGs, which combines multiple colour criteria with the requirement to be located well below the MS. 
Our sample could therefore be slightly biased against galaxies that have been quenched very recently, giving us a somewhat limited view of this so-called post-starburst population.
However, making our own toy model of the evolution of the fraction of recently quenched galaxies, based on the stellar mass functions of \citet{davidzon17}, we predicted a slightly lower recently quenched fraction than \citet{gobat20} at $z_{\mathrm{obs}}\gtrsim0.75$ (by on average $\sim28\%$).
These predictions are in good agreement with our observations at $z_{\mathrm{obs}}<2$, and agree within the uncertainties at $z_{\mathrm{obs}}\gtrsim2-3$, suggesting that our sample is not strongly biased against galaxies that have been quenched very recently.

\subsection{Dust mass fraction}  \label{subsec_analysis_all}

\begin{figure}
\centering
\includegraphics[width=\linewidth]{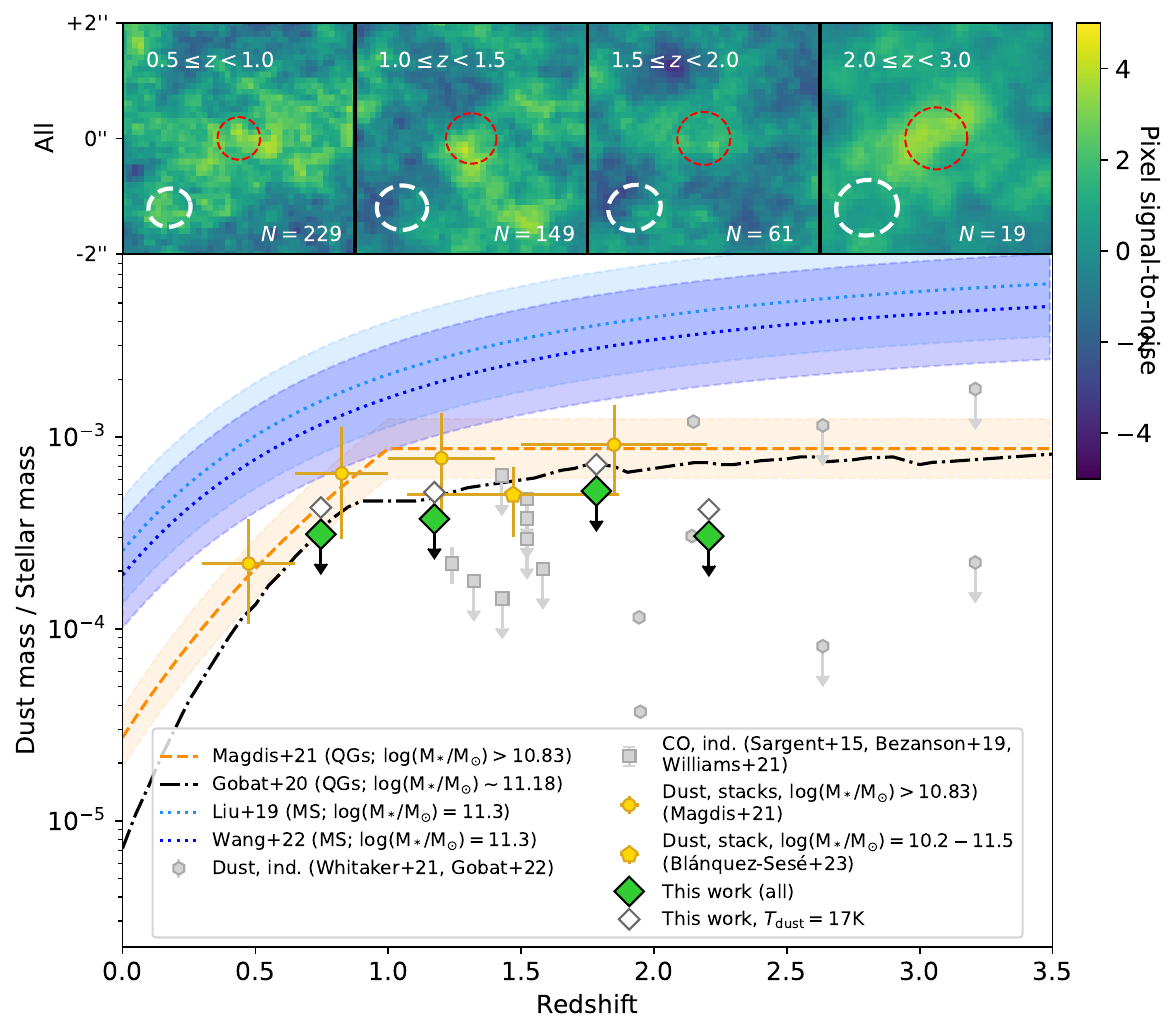}
\caption{Results of our stacking of four different redshift bins: $z_{\mathrm{obs}}=0.5-1$, $1-1.5$, $1.5-2$, and $2-3$. 
\textit{Top panel}: Cutouts of the map centre. 
We show the profile of the synthesised beam as a dashed white shape in the bottom left, and the number of individual galaxies stacked in the bottom right. 
The dashed red circle in the centre shows the aperture used to measure the source emission. 
The colour scaling is set to reflect the signal-to-noise ratio of each pixel. 
\textit{Bottom panel}: Dust-to-stellar mass ratio ($f_{\mathrm{dust}}$) of our stacks (green diamonds) at the weighted mean redshift of each stack in comparison to the $f_{\mathrm{dust}}$ measurements of massive QGs from the literature; grey symbols show individual measurements from dust (grey hexagons) and CO (grey squares), yellow symbols the results from stacking studies, and the dashed yellow and dash-dotted black lines the predictions from data fitting.
The empty diamonds show our upper limits when assuming a lower dust temperature of 17\,K. 
The dashed blue lines and shaded regions show the $f_{\mathrm{dust}}$ evolution of massive MS galaxies from \citet{a3cosmos_2} and \citet{wang22}.}
          \label{fig_fdust_multiplot_onlyall}
\end{figure}

\begin{table}[] \tiny \centering
    \caption[]{Dust mass fraction inferred from our stacks.}   
    \label{tab_results_fdust}
    \begin{tabular}{cccc}
    \hline
    Redshift bin & $\langle z\rangle_w$ & $\log(\langle M_*\rangle_w/M_{\odot})$ & $\langle f_{\mathrm{dust}}\rangle_{\mathrm{pop.}}$  \\
    \hline
    0.5 -- 1.0 & 0.75 & 11.17 & $<3.1\cdot10^{-4}$ \\
    1.0 -- 1.5 & 1.17 & 11.17 & $<3.7\cdot10^{-4}$ \\
    1.5 -- 2.0 & 1.78 & 11.14 & $<5.2\cdot10^{-4}$ \\    
    2.0 -- 3.0 & 2.21 & 11.51 & $<3.0\cdot10^{-4}$ \\
    \hline
    \end{tabular}
\tablefoot{The dust mass fractions are $3\,\sigma$ upper limits.}
\end{table}

We stacked our QGs in the redshift bins of $z_{\mathrm{obs}}=0.5-1$, $1-1.5$, $1.5-2$, and $2-3$, obtaining stringent 3\,$\sigma$ upper limits on the average dust mass fraction (i.e. $M_{\mathrm{dust}}/M_*$) of the QG population $\langle f_{\mathrm{dust}}\rangle_{\mathrm{pop.}}$.
The results are listed in Table ~\ref{tab_results_fdust} and shown in Fig.~\ref{fig_fdust_multiplot_onlyall} as a function of redshift.
Our upper limits yield $\langle f_{\mathrm{dust}}\rangle_{\mathrm{pop.}}\lesssim4\cdot10^{-4}$, which is a factor of $\sim2-7$ below the locus of massive MS galaxies\footnote{The locus of MS galaxies was inferred from their molecular gas mass fraction \citep[][]{a3cosmos_2,wang22} and a GDR of 92 \citep[][]{magdis21}.}, which indicates that the quiescent nature of these galaxies is associated with a low gas content, rather than solely a low star formation efficiency.
These limits are consistent with interferometric measurements of individual massive QGs, in which the dust mass is inferred from either the FIR SED \citep{whitaker21,gobat22} or CO line measurements \citep{sargent15,bezanson19,williams21}. The latter conversion assumes a GDR of 92, which we note is subject to uncertainty \citep[e.g.][]{whitaker21b}.

These individual measurements are, however, in some tension with population-based results from FIR stacking studies, which find $\sim9\cdot10^{-4}$ \citep[][]{gobat18,magdis21}.
To address this issue, \citet{blanquez23} performed an ALMA stacking study, measuring $\langle f_{\mathrm{dust}}\rangle_{\mathrm{pop.}}$ slightly lower than \citet{magdis21}, but agreeing within the errors.
However, \citet{blanquez23} also included galaxies down to stellar masses of $\log(M_*/M_{\odot})=10.2$, so the compared datasets are not consistent. 
Indeed, \citet{blanquez23} support the idea that the gas-to-stellar mass is constant up to $\log(M_*/M_{\odot})\sim11$, but decreases rapidly with higher stellar masses. 
Their $\langle f_{\mathrm{dust}}\rangle_{\mathrm{pop.}}$ is therefore likely biased high compared to samples of more massive QGs.
Our QG selection is consistent with \citet{magdis21} (i.e. $\log(M_*/M_{\odot})>10.8$), yet our measurements are lower by a factor of $\sim2-3$.
This is most likely due to the coarser angular resolution of the (sub-)millimetre images stacked by \citet{magdis21}, who used images from Atacama Submillimeter Telescope Experiment/AzTEC at 1.1\,mm and images from JCMT/Submillimetre Common-User Bolometer Array 2 at 850\,$\mu$m. 
These instruments have an order of magnitude lower angular resolution than ALMA at the same wavelengths and are hence subject to clustering bias, that is, QGs being
located in the same halo as SFGs, and thus not spatially separable in these coarse resolutions.
This can introduce a contamination of the measured fluxes with emission from nearby sources, leading to higher dust mass estimates. Although \citet{magdis21} consider a clustering contribution to their fluxes, such a posterior separation of emission components is much less precise than avoiding confusion effects from the beginning, as is possible with the excellent angular resolution of ALMA.
Naturally, a flux contamination in the measurements of \citet{magdis21} implies that their SED template may also be biased. 
This clustering bias, originating from SFGs with higher dust temperatures than QGs, could indeed lead to greater contamination at shorter than at longer wavelengths. 
Such contamination may shift the peak of the fitted SED, resulting in a slight overestimation of the dust temperature. 
However, this effect would somewhat be mitigated by the higher angular resolution of shorter-wavelength observations, which reduces their sensitivity to clustering bias.
In any case, even if this differential bias implies that QGs have a lower dust temperature than that assumed in \citet{magdis21}, our results remain largely unaffected. 
Indeed, within a reasonable dust temperature range of 17--21\,K \citep[][]{cochrane22}, (i) our template-based rescaling is mostly insensitive to the assumed dust temperature (see Sect.~\ref{subsec_stacking}), and (ii) adopting a lower temperature of 17\,K instead of 21\,K would increase our dust mass estimates by only $\sim\,$27\%, still keeping them below those derived in previous population-based stacking studies.

Our upper limits are only slightly below the trend predicted by the model of \citet{gobat20} at $z<2$. 
This model predicts the evolution of the population-wide gas fraction of QGs based on the QG production rate (see the bottom panel of Fig. \ref{fig_combo}), assuming that QGs start their evolution with a certain initial gas fraction, $f_{0,\mathrm{gas}}$, which decreases exponentially with time: 
\begin{equation}
    f_{\mathrm{gas}}(t_{\mathrm{q}}) = f_{0,\mathrm{gas}}\cdot\exp(-t_{\mathrm{q}}/\tau). \label{eq_tau}
\end{equation}
The model further assumes that $f_{0,\mathrm{gas}}$ is a constant fraction of the gas mass of MS galaxies at the time of quenching, $M_{\mathrm{gas,MS}}(z_{\mathrm{q}},M_{*,\mathrm{q}})$:
$f_{0,\mathrm{gas}}=M_{\mathrm{gas}}(t_{\mathrm{q}}=0)/M_{\mathrm{gas,MS}}(z_{\mathrm{q}},M_{*,\mathrm{q}})$.
The model predicts a nearly constant $\langle f_{\mathrm{dust}}\rangle_{\mathrm{pop.}}$ at $z\gtrsim1$, when the QG population is dominated by recently quenched galaxies, followed by a steep decline of $\langle f_{\mathrm{dust}}\rangle_{\mathrm{pop.}}$ towards $z=0$. 
At $z<2$, where our upper limits are only slightly lower, our fraction of recently quenched galaxies is also largely in agreement with \citet{gobat20} (see Fig.~\ref{fig_combo}).
At these redshifts, the slight discrepancy between our observations and the model may thus simply arise from a modest bias in the model towards higher $f_{0,\mathrm{gas}}$, introduced during the fitting of the \citet{magdis21} data.
At $z>2$, where our upper limit is a factor of $\sim$2.5 lower than the model predictions, this discrepancy likely arises from two factors: (i) our sample at this redshift is somewhat biased against recently quenched galaxies (see Sect.~\ref{subsec_recently_quenched}); and (ii) the model likely overestimates $f_{0,\mathrm{gas}}$ slightly. 
To better assess the relative importance of these two effects, we need to explore the dust content as a function of time since quenching, which we do in the following section.

\subsection{Dust removal} \label{subsec_dustdestruction}

\begin{figure}
\centering
\includegraphics[width=\linewidth]{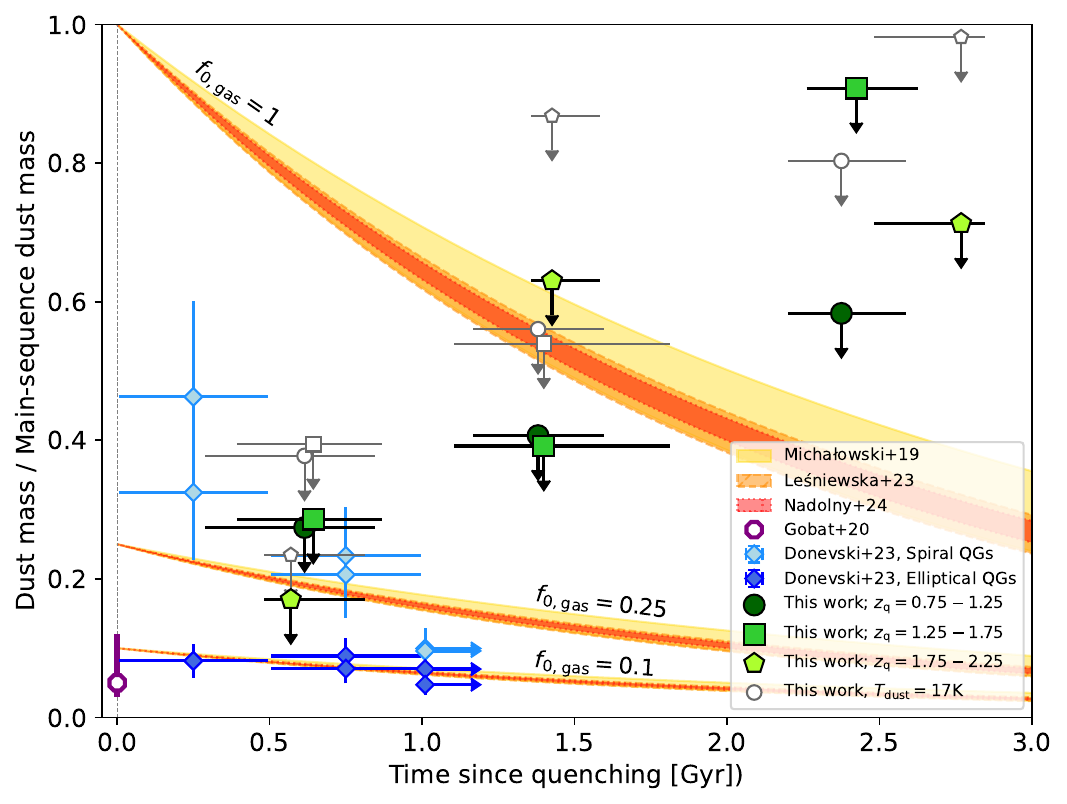}
\caption{Dust mass at the time of observation divided by the dust mass of MS galaxies at the time of quenching as a function of time since quenching for three different bins of quenching redshift (green filled symbols).
The empty symbols show the upper limits when assuming a lower dust temperature of 17\,K.
The yellow, orange-hatched, and red-chequered regions show the dust removal timescales from \citet[][$\tau=2.5\pm0.4$\,Gyr]{michalowski19}, \citet[][$\tau=2.26\pm0.18$\,Gyr]{lesniewska23}, and \citet[][$\tau=2.28\pm0.1$\,Gyr]{nadolny24}, for three different fractional initial gas fractions ($f_{0,\mathrm{gas}}$). 
The empty purple symbol marks the $f_{0,\mathrm{gas}}$ from the fit of \citet{gobat20}.
Light blue and dark blue symbols show dust mass evolution with time since quenching from \citet{donevski23} for spiral and elliptical QGs at $z<0.7$, respectively, where we assume a 30\% error on the dust mass, and the horizontal error bars denote the bin size of \citet{donevski23}.
}
          \label{fig_dustdestruct}
\end{figure}

\begin{table*}[] \small \centering
    \caption[]{Dust mass measured from the stacks.}   
    \label{tab_dust_masses}
    \begin{tabular}{cccccc}
    \hline
    $z_{\mathrm{q}}$ bin& \makecell{Number of\\galaxies} & $\langle t_{\mathrm{q}}\rangle_w$ [Gyr] & $\log(M_{\mathrm{dust}}/M_{\odot})$ & $M_{\mathrm{dust}}/M_{\mathrm{dust,MS}}(z_{\mathrm{q}},M_{*,\mathrm{q}})$ & $\log(\langle M_*\rangle_w/M_{\odot})$\\
    \hline
    0.75 -- 1.25 & 54 & $0.61\substack{+0.23 \\ -0.33}$ & $<7.78$ & $<0.27$ & 11.10\\
                 & 94 & $1.38\substack{+0.22 \\ -0.21}$ & $<7.88$ & $<0.41$ & 11.20\\    
                 & 21 & $2.38\substack{+0.21 \\ -0.18}$ & $<8.02$ & $<0.58$ & 11.25\\    
    \hline
    1.25 -- 1.75 & 89 & $0.64\substack{+0.23 \\ -0.25}$ & $<8.02$ & $<0.29$ & 11.21\\
                 & 63 & $1.40\substack{+0.41 \\ -0.29}$ & $<8.03$ & $<0.39$ & 11.15\\    
                 & 18 & $2.42\substack{+0.20 \\ -0.16}$ & $<8.39$ & $<0.91$ & 11.38\\    
    \hline
    1.75 -- 2.25 & 47 & $0.57\substack{+0.24 \\ -0.09}$ & $<7.93$ & $<0.17$ & 11.19\\
                 & 13 & $1.43\substack{+0.16 \\ -0.07}$ & $<8.37$ & $<0.63$ & 11.21\\    
                 &  6 & $2.77\substack{+0.08 \\ -0.29}$ & $<8.36$ & $<0.71$ & 11.43\\    
    \hline
    \end{tabular}
\tablefoot{The dust masses are $3\,\sigma$ upper limits.}
\end{table*}

Quiescent galaxies are commonly assumed to lose their gas through an initial short quenching process, followed by a longer phase of declining gas and dust content.
As already mentioned, in this framework the parameter $f_{0,\mathrm{gas}}$ expresses the gas mass content right after quenching relative to that on the nominal MS.
The gradual dust removal in QGs after quenching is generally assumed to follow Eq.~\ref{eq_tau} with $\tau\sim2.2-2.5$\,Gyr \citep[e.g.][]{michalowski19,lesniewska23,nadolny24}.
To constrain these two phases of gas removal, we analyse the time evolution of the dust content by (i) grouping galaxies according to their quenching redshift -- thereby isolating systems with a similar initial gas removal phase -- and (ii) grouping them by time since quenching -- capturing galaxies with a consistent after-quenching phase of gas removal.
We consider three bins of quenching redshift, $z_{\mathrm{q}}=0.75-1.25$, $1.25-1.75$, and $1.75-2.25$, and three bins of time since quenching, $t_{\mathrm{q}}<1$\,Gyr, $1-2$\,Gyr, and $2-3$\,Gyr, as shown in the top panel of Fig.~\ref{fig_combo}.
We stacked the QGs in these bins and measured their dust content, obtaining 3\,$\sigma$ upper limits for all stacks.
The measured dust masses of the individual stacks are listed in Table~\ref{tab_dust_masses}.
Figure~\ref{fig_dustdestruct} shows the dust mass as a function of the time since quenching in relation to the dust mass of MS galaxies at the time of quenching (i.e. $M_{\mathrm{dust}}/M_{\mathrm{dust,MS}}(z_{\mathrm{q}},M_{*,\mathrm{q}})$), where the dust mass of MS galaxies was inferred from the MS gas-to-stellar mass relation from \citet{wang22}, converting the molecular gas masses to dust masses using a GDR of 92 \citep[][]{magdis21}.
Given the low star formation after quenching, $r_{\mathrm{SFR}}$ (see Sect. \ref{subsec_cigale_fitting}), we can ignore the stellar mass increase after quenching.

Figure~\ref{fig_dustdestruct} shows that we can place significant upper limits on the dust mass right after quenching ($t_{\mathrm{q}}\lesssim1.5$\,Gyr), whereas at later times, our limits cannot constrain the degenerate parameters $f_{0,\mathrm{gas}}$ and $\tau$ of the \citet{gobat20} model (Eq.~\ref{eq_tau}), although this may be possible in the future. 
Our results show that after $\sim0.6$\,Gyr, the dust mass is decreased by $\gtrsim70\%$.
This is in agreement with the model fitting of \citet{gobat20}, who inferred a value of $f_{0,\mathrm{gas}}=0.05\substack{+0.07\\-0.02}$.
Our results are also mostly consistent with findings by \citet{donevski23} for QGs at $z<0.7$.

In SFGs, the accretion of fresh gas from the circumgalactic medium is crucial to maintain a steady star formation activity, balancing gas consumption \citep[e.g.][]{magnelli20,walter20}. 
If gas accretion continues uninterrupted after quenching, we would expect the gas reservoir of QGs to be replenished within the typical gas consumption timescale of SFGs \citep[$\sim400-700$\,Myr;][]{walter20}.
Therefore, after the initial drop, $M_{\mathrm{dust}}/M_{\mathrm{dust,MS}}$ should increase back to 1 by $t_{\mathrm{q}}\sim1-1.5$\,Gyr if the accreted gas is not pristine and is mostly falling back onto the galaxy from previous outflow episodes, including the one that led to the current quiescence.
However, our upper limits show that $M_{\mathrm{dust}}/M_{\mathrm{dust,MS}}$ is still well below 1 at $t_{\mathrm{q}}=1-2$\,Gyr, and even still at $t_{\mathrm{q}}=2-3$\,Gyr. 
This indicates that, if the dominant quenching mechanism operates via gas removal by outflows, the re-accretion of this dust-rich gas onto the galaxy is also efficiently prevented over long periods of time.

Overall, our results thus support that the bulk of the gas and dust reservoir is removed at the beginning of quiescence, and therefore that the loss of most of the gas is necessary for star formation to cease. 
Larger QG samples will be needed to constrain $f_{0,\mathrm{gas}}$ fully, determine whether it evolves with cosmic time, and place constraints on the gas and dust removal timescale.

\section{Conclusions} \label{sec_summary}

We present an ALMA stacking analysis on a mass-complete sample of 458 QGs with stellar masses $>10^{10.8}\,M_{\odot}$.
We find that most QGs in our sample are quenched between the redshifts $z=1$ and 2, following the peak of cosmic SFR density.
This implies an evolution in the population of massive QGs. 
At $z<1$, massive QGs are dominated by systems that have already been quenched for several gigayears, while at $z>1$, a large fraction of massive QGs are systems that have recently been quenched.
We measured the dust mass fraction as a function of observed redshift and placed stringent upper limits on the average dust mass fraction of the QG population of $\langle f_{\mathrm{dust}}\rangle_{\mathrm{pop.}}\lesssim4\cdot10^{-4}$.
This corroborates the gas and dust measurements obtained via interferometric measurements of single QGs \citep[][]{sargent15,bezanson19,whitaker21,williams21,gobat22}, as opposed to the higher values found by earlier stacking studies \citep[][]{magdis21,blanquez23}.
This implies that quiescence is associated with a very low gas content, more so than a low star formation efficiency.

For the first time, we also measured the dust mass with respect to the dust mass of MS galaxies as a function of the time since quenching. 
We find that the bulk of the gas reservoir of massive QGs ($\gtrsim70\%$) is already removed at the beginning of their quiescent evolution, that is, in the quenching process or within $\sim0.6$\,Gyr afterwards.
These findings favour quenching via a cold gas ejection or consumption scenario acting on short timescales (i.e. a few hundred megayears or less), such as starburst- or AGN-driven outflows \citep[e.g.][]{cicone14}, over a slow gradual quenching through starvation acting on gigayear timescales \citep[e.g.][]{boselli16,trussler20}, or a stabilisation of the cold gas reservoir (i.e. without major gas loss). 
Our results also indicate that, if the dominant quenching mechanism operates via gas removal by outflows, the re-accretion of the ejected dust-rich gas onto the galaxy is efficiently prevented over long periods of time (i.e. on a gigayear timescale).
Finally, our method is also suited to determine the dust removal timescale after quenching, which is currently only inhibited by low number statistics.

With even larger galaxy samples, it will be possible in the future to constrain the dust content of early QGs across cosmic time even more precisely, and to constrain their dust removal timescale.
This will further refine our understanding of quenching mechanisms and timescales. 
The \acosmos{}/\agoodss{} database continues to grow and within a few years will become large enough for our stacks to reach the depth of deep individual QG measurements (i.e. a threefold depth increase, corresponding to a ninefold size increase). 
Future dedicated deep follow-up observations with ALMA on a large sample of as of yet undetected QGs could also greatly advance these efforts.
Additionally, JWST will undoubtedly discover many more QGs at $z>3$, providing the opportunity to study large samples of recently quenched QGs dominating the population at early cosmic times. 

\bibliographystyle{aa}
\bibliography{mybib.bib}

\begin{thebibliography}{92}
\expandafter\ifx\csname natexlab\endcsname\relax\def\natexlab#1{#1}\fi

\bibitem[{{Adscheid} {et~al.}(2024){Adscheid}, {Magnelli}, {Liu}, {Bertoldi},
  {Delvecchio}, {Gruppioni}, {Schinnerer}, {Traina}, {B{\'e}thermin}, \&
  {Gkogkou}}]{adscheid24}
{Adscheid}, S., {Magnelli}, B., {Liu}, D., {et~al.} 2024, \aap, 685, A1

\bibitem[{{Arjona-G{\'a}lvez} {et~al.}(2024){Arjona-G{\'a}lvez}, {Di Cintio},
  \& {Grand}}]{arjona24}
{Arjona-G{\'a}lvez}, E., {Di Cintio}, A., \& {Grand}, R. J.~J. 2024, \aap, 690,
  A286

\bibitem[{{Arnouts} {et~al.}(2002){Arnouts}, {Moscardini}, {Vanzella},
  {Colombi}, {Cristiani}, {Fontana}, {Giallongo}, {Matarrese}, \&
  {Saracco}}]{arnouts02lephare}
{Arnouts}, S., {Moscardini}, L., {Vanzella}, E., {et~al.} 2002, \mnras, 329,
  355

\bibitem[{{Astropy Collaboration} {et~al.}(2022){Astropy Collaboration},
  {Price-Whelan}, {Lim}, {Earl}, {Starkman}, {Bradley}, {Shupe}, {Patil},
  {Corrales}, {Brasseur}, {N{"o}the}, {Donath}, {Tollerud}, {Morris},
  {Ginsburg}, {Vaher}, {Weaver}, {Tocknell}, {Jamieson}, {van Kerkwijk},
  {Robitaille}, {Merry}, {Bachetti}, {G{"u}nther}, {Aldcroft},
  {Alvarado-Montes}, {Archibald}, {B{'o}di}, {Bapat}, {Barentsen}, {Baz{'a}n},
  {Biswas}, {Boquien}, {Burke}, {Cara}, {Cara}, {Conroy}, {Conseil}, {Craig},
  {Cross}, {Cruz}, {D'Eugenio}, {Dencheva}, {Devillepoix}, {Dietrich},
  {Eigenbrot}, {Erben}, {Ferreira}, {Foreman-Mackey}, {Fox}, {Freij}, {Garg},
  {Geda}, {Glattly}, {Gondhalekar}, {Gordon}, {Grant}, {Greenfield}, {Groener},
  {Guest}, {Gurovich}, {Handberg}, {Hart}, {Hatfield-Dodds}, {Homeier},
  {Hosseinzadeh}, {Jenness}, {Jones}, {Joseph}, {Kalmbach}, {Karamehmetoglu},
  {Ka{l}uszy{'n}ski}, {Kelley}, {Kern}, {Kerzendorf}, {Koch}, {Kulumani},
  {Lee}, {Ly}, {Ma}, {MacBride}, {Maljaars}, {Muna}, {Murphy}, {Norman},
  {O'Steen}, {Oman}, {Pacifici}, {Pascual}, {Pascual-Granado}, {Patil},
  {Perren}, {Pickering}, {Rastogi}, {Roulston}, {Ryan}, {Rykoff}, {Sabater},
  {Sakurikar}, {Salgado}, {Sanghi}, {Saunders}, {Savchenko}, {Schwardt},
  {Seifert-Eckert}, {Shih}, {Jain}, {Shukla}, {Sick}, {Simpson},
  {Singanamalla}, {Singer}, {Singhal}, {Sinha}, {Sip{H{o}}cz}, {Spitler},
  {Stansby}, {Streicher}, {{{S}}umak}, {Swinbank}, {Taranu}, {Tewary},
  {Tremblay}, {Val-Borro}, {Van Kooten}, {Vasovi{'c}}, {Verma}, {de Miranda
  Cardoso}, {Williams}, {Wilson}, {Winkel}, {Wood-Vasey}, {Xue}, {Yoachim},
  {Zhang}, {Zonca}, \& {Astropy Project Contributors}}]{astropy22}
{Astropy Collaboration}, {Price-Whelan}, A.~M., {Lim}, P.~L., {et~al.} 2022,
  \apj, 935, 167

\bibitem[{{Astropy Collaboration} {et~al.}(2018){Astropy Collaboration},
  {Price-Whelan}, {Sip{\H{o}}cz}, {G{\"u}nther}, {Lim}, {Crawford}, {Conseil},
  {Shupe}, {Craig}, {Dencheva}, {Ginsburg}, {Vand erPlas}, {Bradley},
  {P{\'e}rez-Su{\'a}rez}, {de Val-Borro}, {Aldcroft}, {Cruz}, {Robitaille},
  {Tollerud}, {Ardelean}, {Babej}, {Bach}, {Bachetti}, {Bakanov}, {Bamford},
  {Barentsen}, {Barmby}, {Baumbach}, {Berry}, {Biscani}, {Boquien}, {Bostroem},
  {Bouma}, {Brammer}, {Bray}, {Breytenbach}, {Buddelmeijer}, {Burke},
  {Calderone}, {Cano Rodr{\'\i}guez}, {Cara}, {Cardoso}, {Cheedella}, {Copin},
  {Corrales}, {Crichton}, {D'Avella}, {Deil}, {Depagne}, {Dietrich}, {Donath},
  {Droettboom}, {Earl}, {Erben}, {Fabbro}, {Ferreira}, {Finethy}, {Fox},
  {Garrison}, {Gibbons}, {Goldstein}, {Gommers}, {Greco}, {Greenfield},
  {Groener}, {Grollier}, {Hagen}, {Hirst}, {Homeier}, {Horton}, {Hosseinzadeh},
  {Hu}, {Hunkeler}, {Ivezi{\'c}}, {Jain}, {Jenness}, {Kanarek}, {Kendrew},
  {Kern}, {Kerzendorf}, {Khvalko}, {King}, {Kirkby}, {Kulkarni}, {Kumar},
  {Lee}, {Lenz}, {Littlefair}, {Ma}, {Macleod}, {Mastropietro}, {McCully},
  {Montagnac}, {Morris}, {Mueller}, {Mumford}, {Muna}, {Murphy}, {Nelson},
  {Nguyen}, {Ninan}, {N{\"o}the}, {Ogaz}, {Oh}, {Parejko}, {Parley}, {Pascual},
  {Patil}, {Patil}, {Plunkett}, {Prochaska}, {Rastogi}, {Reddy Janga},
  {Sabater}, {Sakurikar}, {Seifert}, {Sherbert}, {Sherwood-Taylor}, {Shih},
  {Sick}, {Silbiger}, {Singanamalla}, {Singer}, {Sladen}, {Sooley},
  {Sornarajah}, {Streicher}, {Teuben}, {Thomas}, {Tremblay}, {Turner},
  {Terr{\'o}n}, {van Kerkwijk}, {de la Vega}, {Watkins}, {Weaver}, {Whitmore},
  {Woillez}, {Zabalza}, \& {Astropy Contributors}}]{astropy18}
{Astropy Collaboration}, {Price-Whelan}, A.~M., {Sip{\H{o}}cz}, B.~M., {et~al.}
  2018, \aj, 156, 123

\bibitem[{{Astropy Collaboration} {et~al.}(2013){Astropy Collaboration},
  {Robitaille}, {Tollerud}, {Greenfield}, {Droettboom}, {Bray}, {Aldcroft},
  {Davis}, {Ginsburg}, {Price-Whelan}, {Kerzendorf}, {Conley}, {Crighton},
  {Barbary}, {Muna}, {Ferguson}, {Grollier}, {Parikh}, {Nair}, {Unther},
  {Deil}, {Woillez}, {Conseil}, {Kramer}, {Turner}, {Singer}, {Fox}, {Weaver},
  {Zabalza}, {Edwards}, {Azalee Bostroem}, {Burke}, {Casey}, {Crawford},
  {Dencheva}, {Ely}, {Jenness}, {Labrie}, {Lim}, {Pierfederici}, {Pontzen},
  {Ptak}, {Refsdal}, {Servillat}, \& {Streicher}}]{astropy13}
{Astropy Collaboration}, {Robitaille}, T.~P., {Tollerud}, E.~J., {et~al.} 2013,
  \aap, 558, A33

\bibitem[{{Ball}(1975)}]{ball75}
{Ball}, J.~A. 1975, Methods in Computational Physics, 14, 177

\bibitem[{{Belli} {et~al.}(2024){Belli}, {Park}, {Davies}, {Mendel}, {Johnson},
  {Conroy}, {Benton}, {Bugiani}, {Emami}, {Leja}, {Li}, {Maheson}, {Mathews},
  {Naidu}, {Nelson}, {Tacchella}, {Terrazas}, \& {Weinberger}}]{belli24}
{Belli}, S., {Park}, M., {Davies}, R.~L., {et~al.} 2024, \nat, 630, 54

\bibitem[{{Bendo} {et~al.}(2025){Bendo}, {Bakx}, {Algera}, {Amvrosiadis},
  {Berta}, {Bonavera}, {Cox}, {De Zotti}, {Eales}, {Gonz{\'a}lez-Nuevo},
  {Hagimoto}, {Ismail}, {Riechers}, {Serjeant}, {Smith}, {Temi}, {Tsukui},
  {Urquhart}, \& {Vlahakis}}]{bendo25}
{Bendo}, G.~J., {Bakx}, T.~J.~L.~C., {Algera}, H.~S.~B., {et~al.} 2025, \mnras,
  540, 1560

\bibitem[{{Bezanson} {et~al.}(2019){Bezanson}, {Spilker}, {Williams},
  {Whitaker}, {Narayanan}, {Weiner}, \& {Franx}}]{bezanson19}
{Bezanson}, R., {Spilker}, J., {Williams}, C.~C., {et~al.} 2019, \apjl, 873,
  L19

\bibitem[{{Bl{\'a}nquez-Ses{\'e}} {et~al.}(2023){Bl{\'a}nquez-Ses{\'e}},
  {G{\'o}mez-Guijarro}, {Magdis}, {Magnelli}, {Gobat}, {Daddi}, {Franco},
  {Whitaker}, {Valentino}, {Adscheid}, {Schinnerer}, {Zanella}, {Xiao}, {Wang},
  {Liu}, {Kokorev}, \& {Elbaz}}]{blanquez23}
{Bl{\'a}nquez-Ses{\'e}}, D., {G{\'o}mez-Guijarro}, C., {Magdis}, G.~E.,
  {et~al.} 2023, \aap, 674, A166

\bibitem[{{Boquien} {et~al.}(2019){Boquien}, {Burgarella}, {Roehlly}, {Buat},
  {Ciesla}, {Corre}, {Inoue}, \& {Salas}}]{boquien19cigale}
{Boquien}, M., {Burgarella}, D., {Roehlly}, Y., {et~al.} 2019, \aap, 622, A103

\bibitem[{{Boselli} {et~al.}(2014){Boselli}, {Cortese}, {Boquien}, {Boissier},
  {Catinella}, {Lagos}, \& {Saintonge}}]{boselli14}
{Boselli}, A., {Cortese}, L., {Boquien}, M., {et~al.} 2014, \aap, 564, A66

\bibitem[{{Boselli} {et~al.}(2016){Boselli}, {Roehlly}, {Fossati}, {Buat},
  {Boissier}, {Boquien}, {Burgarella}, {Ciesla}, {Gavazzi}, \&
  {Serra}}]{boselli16}
{Boselli}, A., {Roehlly}, Y., {Fossati}, M., {et~al.} 2016, \aap, 596, A11

\bibitem[{{Brammer} {et~al.}(2008){Brammer}, {van Dokkum}, \&
  {Coppi}}]{brammer08eazy}
{Brammer}, G.~B., {van Dokkum}, P.~G., \& {Coppi}, P. 2008, \apj, 686, 1503

\bibitem[{{Brammer} {et~al.}(2011){Brammer}, {Whitaker}, {van Dokkum},
  {Marchesini}, {Franx}, {Kriek}, {Labb{\'e}}, {Lee}, {Muzzin}, {Quadri},
  {Rudnick}, \& {Williams}}]{brammer11}
{Brammer}, G.~B., {Whitaker}, K.~E., {van Dokkum}, P.~G., {et~al.} 2011, \apj,
  739, 24

\bibitem[{{Bruzual} \& {Charlot}(2003)}]{bruzual03}
{Bruzual}, G. \& {Charlot}, S. 2003, \mnras, 344, 1000

\bibitem[{{Carnall} {et~al.}(2023){Carnall}, {McLure}, {Dunlop}, {McLeod},
  {Wild}, {Cullen}, {Magee}, {Begley}, {Cimatti}, {Donnan}, {Hamadouche},
  {Jewell}, \& {Walker}}]{carnall23}
{Carnall}, A.~C., {McLure}, R.~J., {Dunlop}, J.~S., {et~al.} 2023, \nat, 619,
  716

\bibitem[{{Carniani} {et~al.}(2016){Carniani}, {Marconi}, {Maiolino},
  {Balmaverde}, {Brusa}, {Cano-D{\'\i}az}, {Cicone}, {Comastri}, {Cresci},
  {Fiore}, {Feruglio}, {La Franca}, {Mainieri}, {Mannucci}, {Nagao}, {Netzer},
  {Piconcelli}, {Risaliti}, {Schneider}, \& {Shemmer}}]{carniani16}
{Carniani}, S., {Marconi}, A., {Maiolino}, R., {et~al.} 2016, \aap, 591, A28

\bibitem[{{Charlot} \& {Fall}(2000)}]{charlot00}
{Charlot}, S. \& {Fall}, S.~M. 2000, \apj, 539, 718

\bibitem[{{Cicone} {et~al.}(2014){Cicone}, {Maiolino}, {Sturm},
  {Graci{\'a}-Carpio}, {Feruglio}, {Neri}, {Aalto}, {Davies}, {Fiore},
  {Fischer}, {Garc{\'\i}a-Burillo}, {Gonz{\'a}lez-Alfonso}, {Hailey-Dunsheath},
  {Piconcelli}, \& {Veilleux}}]{cicone14}
{Cicone}, C., {Maiolino}, R., {Sturm}, E., {et~al.} 2014, \aap, 562, A21

\bibitem[{{Ciesla} {et~al.}(2021){Ciesla}, {Buat}, {Boquien}, {Boselli},
  {Elbaz}, \& {Aufort}}]{ciesla21}
{Ciesla}, L., {Buat}, V., {Boquien}, M., {et~al.} 2021, \aap, 653, A6

\bibitem[{{Ciesla} {et~al.}(2017){Ciesla}, {Elbaz}, \& {Fensch}}]{ciesla17}
{Ciesla}, L., {Elbaz}, D., \& {Fensch}, J. 2017, \aap, 608, A41

\bibitem[{{Cochrane} {et~al.}(2022){Cochrane}, {Hayward}, \&
  {Angl{\'e}s-Alc{\'a}zar}}]{cochrane22}
{Cochrane}, R.~K., {Hayward}, C.~C., \& {Angl{\'e}s-Alc{\'a}zar}, D. 2022,
  \apjl, 939, L27

\bibitem[{{Condon}(1997)}]{condon97}
{Condon}, J.~J. 1997, \pasp, 109, 166

\bibitem[{{Conroy} {et~al.}(2015){Conroy}, {van Dokkum}, \&
  {Kravtsov}}]{conroy15}
{Conroy}, C., {van Dokkum}, P.~G., \& {Kravtsov}, A. 2015, \apj, 803, 77

\bibitem[{{Daddi} {et~al.}(2004){Daddi}, {Cimatti}, {Renzini}, {Fontana},
  {Mignoli}, {Pozzetti}, {Tozzi}, \& {Zamorani}}]{daddi04}
{Daddi}, E., {Cimatti}, A., {Renzini}, A., {et~al.} 2004, \apj, 617, 746

\bibitem[{{Dale} {et~al.}(2014){Dale}, {Helou}, {Magdis}, {Armus},
  {D{\'\i}az-Santos}, \& {Shi}}]{dale14}
{Dale}, D.~A., {Helou}, G., {Magdis}, G.~E., {et~al.} 2014, \apj, 784, 83

\bibitem[{{Davidzon} {et~al.}(2017){Davidzon}, {Ilbert}, {Laigle}, {Coupon},
  {McCracken}, {Delvecchio}, {Masters}, {Capak}, {Hsieh}, {Le F{\`e}vre},
  {Tresse}, {Bethermin}, {Chang}, {Faisst}, {Le Floc'h}, {Steinhardt}, {Toft},
  {Aussel}, {Dubois}, {Hasinger}, {Salvato}, {Sanders}, {Scoville}, \&
  {Silverman}}]{davidzon17}
{Davidzon}, I., {Ilbert}, O., {Laigle}, C., {et~al.} 2017, \aap, 605, A70

\bibitem[{{de Graaff} {et~al.}(2024){de Graaff}, {Setton}, {Brammer}, {Cutler},
  {Suess}, {Labbe}, {Leja}, {Weibel}, {Maseda}, {Whitaker}, {Bezanson},
  {Boogaard}, {Cleri}, {De Lucia}, {Franx}, {Greene}, {Hirschmann}, {Matthee},
  {McConachie}, {Naidu}, {Oesch}, {Price}, {Rix}, {Valentino}, {Wang}, \&
  {Williams}}]{degraaf2024}
{de Graaff}, A., {Setton}, D.~J., {Brammer}, G., {et~al.} 2024, arXiv e-prints,
  arXiv:2404.05683

\bibitem[{{Dickinson} {et~al.}(2003){Dickinson}, {Giavalisco}, \& {GOODS
  Team}}]{dickinson03}
{Dickinson}, M., {Giavalisco}, M., \& {GOODS Team}. 2003, in The Mass of
  Galaxies at Low and High Redshift, ed. R.~{Bender} \& A.~{Renzini}, 324

\bibitem[{{Dome} {et~al.}(2024){Dome}, {Tacchella}, {Fialkov}, {Ceverino},
  {Dekel}, {Ginzburg}, {Lapiner}, \& {Looser}}]{dome24}
{Dome}, T., {Tacchella}, S., {Fialkov}, A., {et~al.} 2024, \mnras, 527, 2139

\bibitem[{{Donevski} {et~al.}(2023){Donevski}, {Damjanov}, {Nanni}, {Man},
  {Giulietti}, {Romano}, {Lapi}, {Narayanan}, {Dav{\'e}}, {Shivaei}, {Sohn},
  {Junais}, {Pantoni}, \& {Li}}]{donevski23}
{Donevski}, D., {Damjanov}, I., {Nanni}, A., {et~al.} 2023, \aap, 678, A35

\bibitem[{{Feldmann} \& {Mayer}(2015)}]{feldmann15}
{Feldmann}, R. \& {Mayer}, L. 2015, \mnras, 446, 1939

\bibitem[{{Gehrels}(1986)}]{gehrels86}
{Gehrels}, N. 1986, \apj, 303, 336

\bibitem[{{Gobat} {et~al.}(2018){Gobat}, {Daddi}, {Magdis}, {Bournaud},
  {Sargent}, {Martig}, {Jin}, {Finoguenov}, {B{\'e}thermin}, {Hwang},
  {Renzini}, {Wilson}, {Aretxaga}, {Yun}, {Strazzullo}, \&
  {Valentino}}]{gobat18}
{Gobat}, R., {Daddi}, E., {Magdis}, G., {et~al.} 2018, Nature Astronomy, 2, 239

\bibitem[{{Gobat} {et~al.}(2022){Gobat}, {D'Eugenio}, {Liu}, {Caminha},
  {Daddi}, \& {Bl{\'a}nquez}}]{gobat22}
{Gobat}, R., {D'Eugenio}, C., {Liu}, D., {et~al.} 2022, \aap, 668, L4

\bibitem[{{Gobat} {et~al.}(2020){Gobat}, {Magdis}, {D'Eugenio}, \&
  {Valentino}}]{gobat20}
{Gobat}, R., {Magdis}, G., {D'Eugenio}, C., \& {Valentino}, F. 2020, \aap, 644,
  L7

\bibitem[{{Gonz{\'a}lez-L{\'o}pez} {et~al.}(2019){Gonz{\'a}lez-L{\'o}pez},
  {Decarli}, {Pavesi}, {Walter}, {Aravena}, {Carilli}, {Boogaard}, {Popping},
  {Weiss}, {Assef}, {Bauer}, {Bertoldi}, {Bouwens}, {Contini}, {Cortes}, {Cox},
  {da Cunha}, {Daddi}, {D{\'\i}az-Santos}, {Inami}, {Hodge}, {Ivison}, {Le
  F{\`e}vre}, {Magnelli}, {Oesch}, {Riechers}, {Rix}, {Smail}, {Swinbank},
  {Somerville}, {Uzgil}, \& {van der Werf}}]{gonzales-lopez19}
{Gonz{\'a}lez-L{\'o}pez}, J., {Decarli}, R., {Pavesi}, R., {et~al.} 2019, \apj,
  882, 139

\bibitem[{{Gonz{\'a}lez-L{\'o}pez} {et~al.}(2020){Gonz{\'a}lez-L{\'o}pez},
  {Novak}, {Decarli}, {Walter}, {Aravena}, {Carilli}, {Boogaard}, {Popping},
  {Weiss}, {Assef}, {Bauer}, {Bouwens}, {Cortes}, {Cox}, {Daddi}, {Cunha},
  {D{\'\i}az-Santos}, {Ivison}, {Magnelli}, {Riechers}, {Smail}, {van der
  Werf}, \& {Wagg}}]{gonzales-lopez20}
{Gonz{\'a}lez-L{\'o}pez}, J., {Novak}, M., {Decarli}, R., {et~al.} 2020, \apj,
  897, 91

\bibitem[{{Grogin} {et~al.}(2011){Grogin}, {Kocevski}, {Faber}, {Ferguson},
  {Koekemoer}, {Riess}, {Acquaviva}, {Alexander}, {Almaini}, {Ashby}, {Barden},
  {Bell}, {Bournaud}, {Brown}, {Caputi}, {Casertano}, {Cassata}, {Castellano},
  {Challis}, {Chary}, {Cheung}, {Cirasuolo}, {Conselice}, {Roshan Cooray},
  {Croton}, {Daddi}, {Dahlen}, {Dav{\'e}}, {de Mello}, {Dekel}, {Dickinson},
  {Dolch}, {Donley}, {Dunlop}, {Dutton}, {Elbaz}, {Fazio}, {Filippenko},
  {Finkelstein}, {Fontana}, {Gardner}, {Garnavich}, {Gawiser}, {Giavalisco},
  {Grazian}, {Guo}, {Hathi}, {H{\"a}ussler}, {Hopkins}, {Huang}, {Huang},
  {Jha}, {Kartaltepe}, {Kirshner}, {Koo}, {Lai}, {Lee}, {Li}, {Lotz}, {Lucas},
  {Madau}, {McCarthy}, {McGrath}, {McIntosh}, {McLure}, {Mobasher},
  {Moustakas}, {Mozena}, {Nandra}, {Newman}, {Niemi}, {Noeske}, {Papovich},
  {Pentericci}, {Pope}, {Primack}, {Rajan}, {Ravindranath}, {Reddy}, {Renzini},
  {Rix}, {Robaina}, {Rodney}, {Rosario}, {Rosati}, {Salimbeni}, {Scarlata},
  {Siana}, {Simard}, {Smidt}, {Somerville}, {Spinrad}, {Straughn}, {Strolger},
  {Telford}, {Teplitz}, {Trump}, {van der Wel}, {Villforth}, {Wechsler},
  {Weiner}, {Wiklind}, {Wild}, {Wilson}, {Wuyts}, {Yan}, \& {Yun}}]{grogin11}
{Grogin}, N.~A., {Kocevski}, D.~D., {Faber}, S.~M., {et~al.} 2011, \apjs, 197,
  35

\bibitem[{{Hopkins} {et~al.}(2014){Hopkins}, {Kere{\v{s}}}, {O{\~n}orbe},
  {Faucher-Gigu{\`e}re}, {Quataert}, {Murray}, \& {Bullock}}]{hopkins14}
{Hopkins}, P.~F., {Kere{\v{s}}}, D., {O{\~n}orbe}, J., {et~al.} 2014, \mnras,
  445, 581

\bibitem[{{Ilbert} {et~al.}(2006){Ilbert}, {Arnouts}, {McCracken},
  {Bolzonella}, {Bertin}, {Le F{\`e}vre}, {Mellier}, {Zamorani}, {Pell{\`o}},
  {Iovino}, {Tresse}, {Le Brun}, {Bottini}, {Garilli}, {Maccagni}, {Picat},
  {Scaramella}, {Scodeggio}, {Vettolani}, {Zanichelli}, {Adami}, {Bardelli},
  {Cappi}, {Charlot}, {Ciliegi}, {Contini}, {Cucciati}, {Foucaud}, {Franzetti},
  {Gavignaud}, {Guzzo}, {Marano}, {Marinoni}, {Mazure}, {Meneux}, {Merighi},
  {Paltani}, {Pollo}, {Pozzetti}, {Radovich}, {Zucca}, {Bondi}, {Bongiorno},
  {Busarello}, {de La Torre}, {Gregorini}, {Lamareille}, {Mathez}, {Merluzzi},
  {Ripepi}, {Rizzo}, \& {Vergani}}]{ilbert06lephare}
{Ilbert}, O., {Arnouts}, S., {McCracken}, H.~J., {et~al.} 2006, \aap, 457, 841

\bibitem[{{Ilbert} {et~al.}(2013){Ilbert}, {McCracken}, {Le F{\`e}vre},
  {Capak}, {Dunlop}, {Karim}, {Renzini}, {Caputi}, {Boissier}, {Arnouts},
  {Aussel}, {Comparat}, {Guo}, {Hudelot}, {Kartaltepe}, {Kneib}, {Krogager},
  {Le Floc'h}, {Lilly}, {Mellier}, {Milvang-Jensen}, {Moutard}, {Onodera},
  {Richard}, {Salvato}, {Sanders}, {Scoville}, {Silverman}, {Taniguchi},
  {Tasca}, {Thomas}, {Toft}, {Tresse}, {Vergani}, {Wolk}, \& {Zirm}}]{ilbert13}
{Ilbert}, O., {McCracken}, H.~J., {Le F{\`e}vre}, O., {et~al.} 2013, \aap, 556,
  A55

\bibitem[{{IRAC Instrument Team} \& {IRAC Instrument Support
  Team}(2021)}]{irac}
{IRAC Instrument Team} \& {IRAC Instrument Support Team}. 2021, {IRAC
  Instrument Handbook}, NASA IPAC DataSet, IRSA486

\bibitem[{{Jin} {et~al.}(2018){Jin}, {Daddi}, {Liu}, {Smol{\v{c}}i{\'c}},
  {Schinnerer}, {Calabr{\`o}}, {Gu}, {Delhaize}, {Delvecchio}, {Gao},
  {Salvato}, {Puglisi}, {Dickinson}, {Bertoldi}, {Sargent}, {Novak}, {Magdis},
  {Aretxaga}, {Wilson}, \& {Capak}}]{jin18}
{Jin}, S., {Daddi}, E., {Liu}, D., {et~al.} 2018, \apj, 864, 56

\bibitem[{{Koprowski} {et~al.}(2024){Koprowski}, {Wijesekera}, {Dunlop},
  {McLeod}, {Micha{\l}owski}, {Lisiecki}, \& {McLure}}]{koprowski24}
{Koprowski}, M.~P., {Wijesekera}, J.~V., {Dunlop}, J.~S., {et~al.} 2024, \aap,
  691, A164

\bibitem[{{Kriek} {et~al.}(2009){Kriek}, {van Dokkum}, {Labb{\'e}}, {Franx},
  {Illingworth}, {Marchesini}, \& {Quadri}}]{kriek09fast}
{Kriek}, M., {van Dokkum}, P.~G., {Labb{\'e}}, I., {et~al.} 2009, \apj, 700,
  221

\bibitem[{{Larson} {et~al.}(1980){Larson}, {Tinsley}, \& {Caldwell}}]{larson80}
{Larson}, R.~B., {Tinsley}, B.~M., \& {Caldwell}, C.~N. 1980, \apj, 237, 692

\bibitem[{{Le{\'s}niewska} {et~al.}(2023){Le{\'s}niewska}, {Micha{\l}owski},
  {Gall}, {Hjorth}, {Nadolny}, {Ryzhov}, \& {Solar}}]{lesniewska23}
{Le{\'s}niewska}, A., {Micha{\l}owski}, M.~J., {Gall}, C., {et~al.} 2023, \apj,
  953, 27

\bibitem[{{Li} {et~al.}(2020){Li}, {Li}, {Bryan}, {Ostriker}, \&
  {Quataert}}]{li20}
{Li}, M., {Li}, Y., {Bryan}, G.~L., {Ostriker}, E.~C., \& {Quataert}, E. 2020,
  \apj, 898, 23

\bibitem[{{Lin} {et~al.}(2019){Lin}, {Hsieh}, {Pan}, {Rembold}, {S{\'a}nchez},
  {Argudo-Fern{\'a}ndez}, {Rowlands}, {Belfiore}, {Bizyaev}, {Lacerna},
  {Riffel}, {Rong}, {Yuan}, {Drory}, {Maiolino}, \& {Wilcots}}]{lin19}
{Lin}, L., {Hsieh}, B.-C., {Pan}, H.-A., {et~al.} 2019, \apj, 872, 50

\bibitem[{{Liu} {et~al.}(2019{\natexlab{a}}){Liu}, {Lang}, {Magnelli},
  {Schinnerer}, {Leslie}, {Fudamoto}, {Bondi}, {Groves}, {Jim{\'e}nez-Andrade},
  {Harrington}, {Karim}, {Oesch}, {Sargent}, {Vardoulaki}, {B{\v{a}}descu},
  {Moser}, {Bertoldi}, {Battisti}, {da Cunha}, {Zavala}, {Vaccari}, {Davidzon},
  {Riechers}, \& {Aravena}}]{a3cosmos_1}
{Liu}, D., {Lang}, P., {Magnelli}, B., {et~al.} 2019{\natexlab{a}}, \apjs, 244,
  40

\bibitem[{{Liu} {et~al.}(2019{\natexlab{b}}){Liu}, {Schinnerer}, {Groves},
  {Magnelli}, {Lang}, {Leslie}, {Jim{\'e}nez-Andrade}, {Riechers}, {Popping},
  {Magdis}, {Daddi}, {Sargent}, {Gao}, {Fudamoto}, {Oesch}, \&
  {Bertoldi}}]{a3cosmos_2}
{Liu}, D., {Schinnerer}, E., {Groves}, B., {et~al.} 2019{\natexlab{b}}, \apj,
  887, 235

\bibitem[{{Madau} \& {Dickinson}(2014)}]{madau14}
{Madau}, P. \& {Dickinson}, M. 2014, \araa, 52, 415

\bibitem[{{Magdis} {et~al.}(2012){Magdis}, {Daddi}, {B{\'e}thermin}, {Sargent},
  {Elbaz}, {Pannella}, {Dickinson}, {Dannerbauer}, {da Cunha}, {Walter},
  {Rigopoulou}, {Charmandaris}, {Hwang}, \& {Kartaltepe}}]{magdis12}
{Magdis}, G.~E., {Daddi}, E., {B{\'e}thermin}, M., {et~al.} 2012, \apj, 760, 6

\bibitem[{{Magdis} {et~al.}(2021){Magdis}, {Gobat}, {Valentino}, {Daddi},
  {Zanella}, {Kokorev}, {Toft}, {Jin}, \& {Whitaker}}]{magdis21}
{Magdis}, G.~E., {Gobat}, R., {Valentino}, F., {et~al.} 2021, \aap, 647, A33

\bibitem[{{Magnelli} {et~al.}(2024){Magnelli}, {Adscheid}, {Wang}, {Ciesla},
  {Daddi}, {Delvecchio}, {Elbaz}, {Fudamoto}, {Fukushima}, {Franco},
  {G{\'o}mez-Guijarro}, {Gruppioni}, {Jim{\'e}nez-Andrade}, {Liu}, {Oesch},
  {Schinnerer}, \& {Traina}}]{magnelli24}
{Magnelli}, B., {Adscheid}, S., {Wang}, T.-M., {et~al.} 2024, \aap, 688, A55

\bibitem[{{Magnelli} {et~al.}(2020){Magnelli}, {Boogaard}, {Decarli},
  {G{\'o}nzalez-L{\'o}pez}, {Novak}, {Popping}, {Smail}, {Walter}, {Aravena},
  {Assef}, {Bauer}, {Bertoldi}, {Carilli}, {Cortes}, {Cunha}, {Daddi},
  {D{\'\i}az-Santos}, {Inami}, {Ivison}, {F{\`e}vre}, {Oesch}, {Riechers},
  {Rix}, {Sargent}, {Werf}, {Wagg}, \& {Weiss}}]{magnelli20}
{Magnelli}, B., {Boogaard}, L., {Decarli}, R., {et~al.} 2020, \apj, 892, 66

\bibitem[{{Magnelli} {et~al.}(2013){Magnelli}, {Popesso}, {Berta}, {Pozzi},
  {Elbaz}, {Lutz}, {Dickinson}, {Altieri}, {Andreani}, {Aussel},
  {B{\'e}thermin}, {Bongiovanni}, {Cepa}, {Charmandaris}, {Chary}, {Cimatti},
  {Daddi}, {F{\"o}rster Schreiber}, {Genzel}, {Gruppioni}, {Harwit}, {Hwang},
  {Ivison}, {Magdis}, {Maiolino}, {Murphy}, {Nordon}, {Pannella}, {P{\'e}rez
  Garc{\'\i}a}, {Poglitsch}, {Rosario}, {Sanchez-Portal}, {Santini}, {Scott},
  {Sturm}, {Tacconi}, \& {Valtchanov}}]{magnelli13}
{Magnelli}, B., {Popesso}, P., {Berta}, S., {et~al.} 2013, \aap, 553, A132

\bibitem[{{Martig} {et~al.}(2009){Martig}, {Bournaud}, {Teyssier}, \&
  {Dekel}}]{martig09}
{Martig}, M., {Bournaud}, F., {Teyssier}, R., \& {Dekel}, A. 2009, \apj, 707,
  250

\bibitem[{{McMullin} {et~al.}(2007){McMullin}, {Waters}, {Schiebel}, {Young},
  \& {Golap}}]{casa}
{McMullin}, J.~P., {Waters}, B., {Schiebel}, D., {Young}, W., \& {Golap}, K.
  2007, in Astronomical Society of the Pacific Conference Series, Vol. 376,
  Astronomical Data Analysis Software and Systems XVI, ed. R.~A. {Shaw},
  F.~{Hill}, \& D.~J. {Bell}, 127

\bibitem[{{Micha{\l}owski} {et~al.}(2024){Micha{\l}owski}, {Gall}, {Hjorth},
  {Frayer}, {Tsai}, {Rowlands}, {Takeuchi}, {Le{\'s}niewska}, {Behrendt},
  {Bourne}, {Hughes}, {Koprowski}, {Nadolny}, {Ryzhov}, {Solar}, {Spring},
  {Zavala}, \& {Bartczak}}]{michalowski24}
{Micha{\l}owski}, M.~J., {Gall}, C., {Hjorth}, J., {et~al.} 2024, \apj, 964,
  129

\bibitem[{{Micha{\l}owski} {et~al.}(2019){Micha{\l}owski}, {Hjorth}, {Gall},
  {Frayer}, {Tsai}, {Hirashita}, {Rowlands}, {Takeuchi}, {Le{\'s}niewska},
  {Behrendt}, {Bourne}, {Hughes}, {Spring}, {Zavala}, \&
  {Bartczak}}]{michalowski19}
{Micha{\l}owski}, M.~J., {Hjorth}, J., {Gall}, C., {et~al.} 2019, \aap, 632,
  A43

\bibitem[{{Nadolny} {et~al.}(2024){Nadolny}, {Micha{\l}owski}, {Parente},
  {Hjorth}, {Gall}, {Le{\'s}niewska}, {Solar}, {Nowaczyk}, \&
  {Ryzhov}}]{nadolny24}
{Nadolny}, J., {Micha{\l}owski}, M.~J., {Parente}, M., {et~al.} 2024, \aap,
  689, A210

\bibitem[{{Noeske} {et~al.}(2007){Noeske}, {Weiner}, {Faber}, {Papovich},
  {Koo}, {Somerville}, {Bundy}, {Conselice}, {Newman}, {Schiminovich}, {Le
  Floc'h}, {Coil}, {Rieke}, {Lotz}, {Primack}, {Barmby}, {Cooper}, {Davis},
  {Ellis}, {Fazio}, {Guhathakurta}, {Huang}, {Kassin}, {Martin}, {Phillips},
  {Rich}, {Small}, {Willmer}, \& {Wilson}}]{noeske07}
{Noeske}, K.~G., {Weiner}, B.~J., {Faber}, S.~M., {et~al.} 2007, \apjl, 660,
  L43

\bibitem[{{Page} {et~al.}(2012){Page}, {Symeonidis}, {Vieira}, {Altieri},
  {Amblard}, {Arumugam}, {Aussel}, {Babbedge}, {Blain}, {Bock}, {Boselli},
  {Buat}, {Castro-Rodr{\'\i}guez}, {Cava}, {Chanial}, {Clements}, {Conley},
  {Conversi}, {Cooray}, {Dowell}, {Dubois}, {Dunlop}, {Dwek}, {Dye}, {Eales},
  {Elbaz}, {Farrah}, {Fox}, {Franceschini}, {Gear}, {Glenn}, {Griffin},
  {Halpern}, {Hatziminaoglou}, {Ibar}, {Isaak}, {Ivison}, {Lagache},
  {Levenson}, {Lu}, {Madden}, {Maffei}, {Mainetti}, {Marchetti}, {Nguyen},
  {O'Halloran}, {Oliver}, {Omont}, {Panuzzo}, {Papageorgiou}, {Pearson},
  {P{\'e}rez-Fournon}, {Pohlen}, {Rawlings}, {Rigopoulou}, {Riguccini},
  {Rizzo}, {Rodighiero}, {Roseboom}, {Rowan-Robinson}, {Portal}, {Schulz},
  {Scott}, {Seymour}, {Shupe}, {Smith}, {Stevens}, {Trichas}, {Tugwell},
  {Vaccari}, {Valtchanov}, {Viero}, {Vigroux}, {Wang}, {Ward}, {Wright}, {Xu},
  \& {Zemcov}}]{page12}
{Page}, M.~J., {Symeonidis}, M., {Vieira}, J.~D., {et~al.} 2012, \nat, 485, 213

\bibitem[{{Planck Collaboration} {et~al.}(2011){Planck Collaboration},
  {Abergel}, {Ade}, {Aghanim}, {Arnaud}, {Ashdown}, {Aumont}, {Baccigalupi},
  {Balbi}, {Banday}, {Barreiro}, {Bartlett}, {Battaner}, {Benabed},
  {Beno{\^\i}t}, {Bernard}, {Bersanelli}, {Bhatia}, {Bock}, {Bonaldi}, {Bond},
  {Borrill}, {Bouchet}, {Boulanger}, {Bucher}, {Burigana}, {Cabella},
  {Cardoso}, {Catalano}, {Cay{\'o}n}, {Challinor}, {Chamballu}, {Chiang},
  {Chiang}, {Christensen}, {Clements}, {Colombi}, {Couchot}, {Coulais},
  {Crill}, {Cuttaia}, {Danese}, {Davies}, {Davis}, {de Bernardis}, {de
  Gasperis}, {de Rosa}, {de Zotti}, {Delabrouille}, {Delouis}, {D{\'e}sert},
  {Dickinson}, {Dobashi}, {Donzelli}, {Dor{\'e}}, {D{\"o}rl}, {Douspis},
  {Dupac}, {Efstathiou}, {En{\ss}lin}, {Eriksen}, {Finelli}, {Forni},
  {Frailis}, {Franceschi}, {Galeotta}, {Ganga}, {Giard}, {Giardino},
  {Giraud-H{\'e}raud}, {Gonz{\'a}lez-Nuevo}, {G{\'o}rski}, {Gratton},
  {Gregorio}, {Gruppuso}, {Guillet}, {Hansen}, {Harrison},
  {Henrot-Versill{\'e}}, {Herranz}, {Hildebrandt}, {Hivon}, {Hobson}, {Holmes},
  {Hovest}, {Hoyland}, {Huffenberger}, {Jaffe}, {Jones}, {Jones}, {Juvela},
  {Keih{\"a}nen}, {Keskitalo}, {Kisner}, {Kneissl}, {Knox}, {Kurki-Suonio},
  {Lagache}, {Lamarre}, {Lasenby}, {Laureijs}, {Lawrence}, {Leach}, {Leonardi},
  {Leroy}, {Linden-V{\o}rnle}, {L{\'o}pez-Caniego}, {Lubin},
  {Mac{\'\i}as-P{\'e}rez}, {MacTavish}, {Maffei}, {Mandolesi}, {Mann}, {Maris},
  {Marshall}, {Martin}, {Mart{\'\i}nez-Gonz{\'a}lez}, {Masi}, {Matarrese},
  {Matthai}, {Mazzotta}, {McGehee}, {Meinhold}, {Melchiorri}, {Mendes},
  {Mennella}, {Mitra}, {Miville-Desch{\^e}nes}, {Moneti}, {Montier},
  {Morgante}, {Mortlock}, {Munshi}, {Murphy}, {Naselsky}, {Natoli},
  {Netterfield}, {N{\o}rgaard-Nielsen}, {Noviello}, {Novikov}, {Novikov},
  {Osborne}, {Pajot}, {Paladini}, {Pasian}, {Patanchon}, {Perdereau},
  {Perotto}, {Perrotta}, {Piacentini}, {Piat}, {Plaszczynski}, {Pointecouteau},
  {Polenta}, {Ponthieu}, {Poutanen}, {Pr{\'e}zeau}, {Prunet}, {Puget}, {Reach},
  {Rebolo}, {Reinecke}, {Renault}, {Ricciardi}, {Riller}, {Ristorcelli},
  {Rocha}, {Rosset}, {Rubi{\~n}o-Mart{\'\i}n}, {Rusholme}, {Sandri}, {Santos},
  {Savini}, {Scott}, {Seiffert}, {Shellard}, {Smoot}, {Starck}, {Stivoli},
  {Stolyarov}, {Sudiwala}, {Sygnet}, {Tauber}, {Terenzi}, {Toffolatti},
  {Tomasi}, {Torre}, {Tristram}, {Tuovinen}, {Umana}, {Valenziano},
  {Verstraete}, {Vielva}, {Villa}, {Vittorio}, {Wade}, {Wandelt}, {Yvon},
  {Zacchei}, \& {Zonca}}]{planck11}
{Planck Collaboration}, {Abergel}, A., {Ade}, P.~A.~R., {et~al.} 2011, \aap,
  536, A25

\bibitem[{{Popesso} {et~al.}(2023){Popesso}, {Concas}, {Cresci}, {Belli},
  {Rodighiero}, {Inami}, {Dickinson}, {Ilbert}, {Pannella}, \&
  {Elbaz}}]{popesso23}
{Popesso}, P., {Concas}, A., {Cresci}, G., {et~al.} 2023, \mnras, 519, 1526

\bibitem[{{Salim}(2014)}]{salim14}
{Salim}, S. 2014, Serbian Astronomical Journal, 189, 1

\bibitem[{{Salpeter}(1955)}]{salpeter55}
{Salpeter}, E.~E. 1955, \apj, 121, 161

\bibitem[{{Sargent} {et~al.}(2015){Sargent}, {Daddi}, {Bournaud}, {Onodera},
  {Feruglio}, {Martig}, {Gobat}, {Dannerbauer}, \& {Schinnerer}}]{sargent15}
{Sargent}, M.~T., {Daddi}, E., {Bournaud}, F., {et~al.} 2015, \apjl, 806, L20

\bibitem[{{Schreiber} {et~al.}(2018){Schreiber}, {Elbaz}, {Pannella}, {Ciesla},
  {Wang}, \& {Franco}}]{schreiber18}
{Schreiber}, C., {Elbaz}, D., {Pannella}, M., {et~al.} 2018, \aap, 609, A30

\bibitem[{{Schreiber} {et~al.}(2015){Schreiber}, {Pannella}, {Elbaz},
  {B{\'e}thermin}, {Inami}, {Dickinson}, {Magnelli}, {Wang}, {Aussel}, {Daddi},
  {Juneau}, {Shu}, {Sargent}, {Buat}, {Faber}, {Ferguson}, {Giavalisco},
  {Koekemoer}, {Magdis}, {Morrison}, {Papovich}, {Santini}, \&
  {Scott}}]{schreiber15}
{Schreiber}, C., {Pannella}, M., {Elbaz}, D., {et~al.} 2015, \aap, 575, A74

\bibitem[{{Scoville} {et~al.}(2007){Scoville}, {Aussel}, {Brusa}, {Capak},
  {Carollo}, {Elvis}, {Giavalisco}, {Guzzo}, {Hasinger}, {Impey}, {Kneib},
  {LeFevre}, {Lilly}, {Mobasher}, {Renzini}, {Rich}, {Sanders}, {Schinnerer},
  {Schminovich}, {Shopbell}, {Taniguchi}, \& {Tyson}}]{cosmos}
{Scoville}, N., {Aussel}, H., {Brusa}, M., {et~al.} 2007, \apjs, 172, 1

\bibitem[{{Shu} {et~al.}(1987){Shu}, {Adams}, \& {Lizano}}]{shu87}
{Shu}, F.~H., {Adams}, F.~C., \& {Lizano}, S. 1987, \araa, 25, 23

\bibitem[{{Speagle} {et~al.}(2014){Speagle}, {Steinhardt}, {Capak}, \&
  {Silverman}}]{speagle14}
{Speagle}, J.~S., {Steinhardt}, C.~L., {Capak}, P.~L., \& {Silverman}, J.~D.
  2014, \apjs, 214, 15

\bibitem[{{Straatman} {et~al.}(2016){Straatman}, {Spitler}, {Quadri},
  {Labb{\'e}}, {Glazebrook}, {Persson}, {Papovich}, {Tran}, {Brammer},
  {Cowley}, {Tomczak}, {Nanayakkara}, {Alcorn}, {Allen}, {Broussard}, {van
  Dokkum}, {Forrest}, {van Houdt}, {Kacprzak}, {Kawinwanichakij}, {Kelson},
  {Lee}, {McCarthy}, {Mehrtens}, {Monson}, {Murphy}, {Rees}, {Tilvi}, \&
  {Whitaker}}]{straatman16}
{Straatman}, C. M.~S., {Spitler}, L.~R., {Quadri}, R.~F., {et~al.} 2016, \apj,
  830, 51

\bibitem[{{Trussler} {et~al.}(2020){Trussler}, {Maiolino}, {Maraston}, {Peng},
  {Thomas}, {Goddard}, \& {Lian}}]{trussler20}
{Trussler}, J., {Maiolino}, R., {Maraston}, C., {et~al.} 2020, \mnras, 491,
  5406

\bibitem[{{Valentino} {et~al.}(2023){Valentino}, {Brammer}, {Gould}, {Kokorev},
  {Fujimoto}, {Jespersen}, {Vijayan}, {Weaver}, {Ito}, {Tanaka}, {Ilbert},
  {Magdis}, {Whitaker}, {Faisst}, {Gallazzi}, {Gillman}, {Gim{\'e}nez-Arteaga},
  {G{\'o}mez-Guijarro}, {Kubo}, {Heintz}, {Hirschmann}, {Oesch}, {Onodera},
  {Rizzo}, {Lee}, {Strait}, \& {Toft}}]{valentino23}
{Valentino}, F., {Brammer}, G., {Gould}, K. M.~L., {et~al.} 2023, \apj, 947, 20

\bibitem[{{Walter} {et~al.}(2020){Walter}, {Carilli}, {Neeleman}, {Decarli},
  {Popping}, {Somerville}, {Aravena}, {Bertoldi}, {Boogaard}, {Cox}, {da
  Cunha}, {Magnelli}, {Obreschkow}, {Riechers}, {Rix}, {Smail}, {Weiss},
  {Assef}, {Bauer}, {Bouwens}, {Contini}, {Cortes}, {Daddi}, {Diaz-Santos},
  {Gonz{\'a}lez-L{\'o}pez}, {Hennawi}, {Hodge}, {Inami}, {Ivison}, {Oesch},
  {Sargent}, {van der Werf}, {Wagg}, \& {Yung}}]{walter20}
{Walter}, F., {Carilli}, C., {Neeleman}, M., {et~al.} 2020, \apj, 902, 111

\bibitem[{{Wang} {et~al.}(2024){Wang}, {Magnelli}, {Schinnerer}, {Liu},
  {Jim{\'e}nez-Andrade}, {Karoumpis}, {Adscheid}, \& {Bertoldi}}]{wang24}
{Wang}, T.-M., {Magnelli}, B., {Schinnerer}, E., {et~al.} 2024, \aap, 681, A110

\bibitem[{{Wang} {et~al.}(2022){Wang}, {Magnelli}, {Schinnerer}, {Liu},
  {Modak}, {Jim{\'e}nez-Andrade}, {Karoumpis}, {Kokorev}, \&
  {Bertoldi}}]{wang22}
{Wang}, T.-M., {Magnelli}, B., {Schinnerer}, E., {et~al.} 2022, \aap, 660, A142

\bibitem[{{Weaver} {et~al.}(2023){Weaver}, {Davidzon}, {Toft}, {Ilbert},
  {McCracken}, {Gould}, {Jespersen}, {Steinhardt}, {Lagos}, {Capak}, {Casey},
  {Chartab}, {Faisst}, {Hayward}, {Kartaltepe}, {Kauffmann}, {Koekemoer},
  {Kokorev}, {Laigle}, {Liu}, {Long}, {Magdis}, {McPartland}, {Milvang-Jensen},
  {Mobasher}, {Moneti}, {Peng}, {Sanders}, {Shuntov}, {Sneppen}, {Valentino},
  {Zalesky}, \& {Zamorani}}]{weaver23}
{Weaver}, J.~R., {Davidzon}, I., {Toft}, S., {et~al.} 2023, \aap, 677, A184

\bibitem[{{Weaver} {et~al.}(2022){Weaver}, {Kauffmann}, {Ilbert}, {McCracken},
  {Moneti}, {Toft}, {Brammer}, {Shuntov}, {Davidzon}, {Hsieh}, {Laigle},
  {Anastasiou}, {Jespersen}, {Vinther}, {Capak}, {Casey}, {McPartland},
  {Milvang-Jensen}, {Mobasher}, {Sanders}, {Zalesky}, {Arnouts}, {Aussel},
  {Dunlop}, {Faisst}, {Franx}, {Furtak}, {Fynbo}, {Gould}, {Greve}, {Gwyn},
  {Kartaltepe}, {Kashino}, {Koekemoer}, {Kokorev}, {Le F{\`e}vre}, {Lilly},
  {Masters}, {Magdis}, {Mehta}, {Peng}, {Riechers}, {Salvato}, {Sawicki},
  {Scarlata}, {Scoville}, {Shirley}, {Silverman}, {Sneppen}, {Smolc̆i{\'c}},
  {Steinhardt}, {Stern}, {Tanaka}, {Taniguchi}, {Teplitz}, {Vaccari}, {Wang},
  \& {Zamorani}}]{cosmos2020}
{Weaver}, J.~R., {Kauffmann}, O.~B., {Ilbert}, O., {et~al.} 2022, \apjs, 258,
  11

\bibitem[{{Weibel} {et~al.}(2024){Weibel}, {de Graaff}, {Setton}, {Miller},
  {Oesch}, {Brammer}, {Lagos}, {Whitaker}, {Williams}, {Baggen}, {Bezanson},
  {Boogaard}, {Cleri}, {Greene}, {Hirschmann}, {Hviding}, {Kuruvanthodi},
  {Labb{\'e}}, {Leja}, {Maseda}, {Matthee}, {McConachie}, {Naidu},
  {Roberts-Borsani}, {Schaerer}, {Suess}, {Valentino}, {van Dokkum}, \&
  {Wang}}]{weibel24}
{Weibel}, A., {de Graaff}, A., {Setton}, D.~J., {et~al.} 2024, arXiv e-prints,
  arXiv:2409.03829

\bibitem[{{Whitaker} {et~al.}(2021{\natexlab{a}}){Whitaker}, {Narayanan},
  {Williams}, {Li}, {Spilker}, {Dav{\'e}}, {Akhshik}, {Akins}, {Bezanson},
  {Katz}, {Leja}, {Magdis}, {Mowla}, {Nelson}, {Pope}, {Privon}, {Toft}, \&
  {Valentino}}]{whitaker21b}
{Whitaker}, K.~E., {Narayanan}, D., {Williams}, C.~C., {et~al.}
  2021{\natexlab{a}}, \apjl, 922, L30

\bibitem[{{Whitaker} {et~al.}(2021{\natexlab{b}}){Whitaker}, {Williams},
  {Mowla}, {Spilker}, {Toft}, {Narayanan}, {Pope}, {Magdis}, {van Dokkum},
  {Akhshik}, {Bezanson}, {Brammer}, {Leja}, {Man}, {Nelson}, {Richard},
  {Pacifici}, {Sharon}, \& {Valentino}}]{whitaker21}
{Whitaker}, K.~E., {Williams}, C.~C., {Mowla}, L., {et~al.} 2021{\natexlab{b}},
  \nat, 597, 485

\bibitem[{{Williams} {et~al.}(2021){Williams}, {Spilker}, {Whitaker},
  {Dav{\'e}}, {Woodrum}, {Brammer}, {Bezanson}, {Narayanan}, \&
  {Weiner}}]{williams21}
{Williams}, C.~C., {Spilker}, J.~S., {Whitaker}, K.~E., {et~al.} 2021, \apj,
  908, 54

\bibitem[{{Williams} {et~al.}(2009){Williams}, {Quadri}, {Franx}, {van Dokkum},
  \& {Labb{\'e}}}]{williams09}
{Williams}, R.~J., {Quadri}, R.~F., {Franx}, M., {van Dokkum}, P., \&
  {Labb{\'e}}, I. 2009, \apj, 691, 1879

\bibitem[{{Wuyts} {et~al.}(2011){Wuyts}, {F{\"o}rster Schreiber}, {van der
  Wel}, {Magnelli}, {Guo}, {Genzel}, {Lutz}, {Aussel}, {Barro}, {Berta},
  {Cava}, {Graci{\'a}-Carpio}, {Hathi}, {Huang}, {Kocevski}, {Koekemoer},
  {Lee}, {Le Floc'h}, {McGrath}, {Nordon}, {Popesso}, {Pozzi}, {Riguccini},
  {Rodighiero}, {Saintonge}, \& {Tacconi}}]{wuyts11}
{Wuyts}, S., {F{\"o}rster Schreiber}, N.~M., {van der Wel}, A., {et~al.} 2011,
  \apj, 742, 96

\bibitem[{{Young} {et~al.}(2011){Young}, {Bureau}, {Davis}, {Combes},
  {McDermid}, {Alatalo}, {Blitz}, {Bois}, {Bournaud}, {Cappellari}, {Davies},
  {de Zeeuw}, {Emsellem}, {Khochfar}, {Krajnovi{\'c}}, {Kuntschner},
  {Lablanche}, {Morganti}, {Naab}, {Oosterloo}, {Sarzi}, {Scott}, {Serra}, \&
  {Weijmans}}]{young11}
{Young}, L.~M., {Bureau}, M., {Davis}, T.~A., {et~al.} 2011, \mnras, 414, 940

\end{thebibliography}

\begin{appendix}
\onecolumn

\section{\texttt{CIGALE} mock analysis} \label{app_cigale_mock}

\begin{figure*}[h]
\centering
\includegraphics[width=0.9\linewidth]{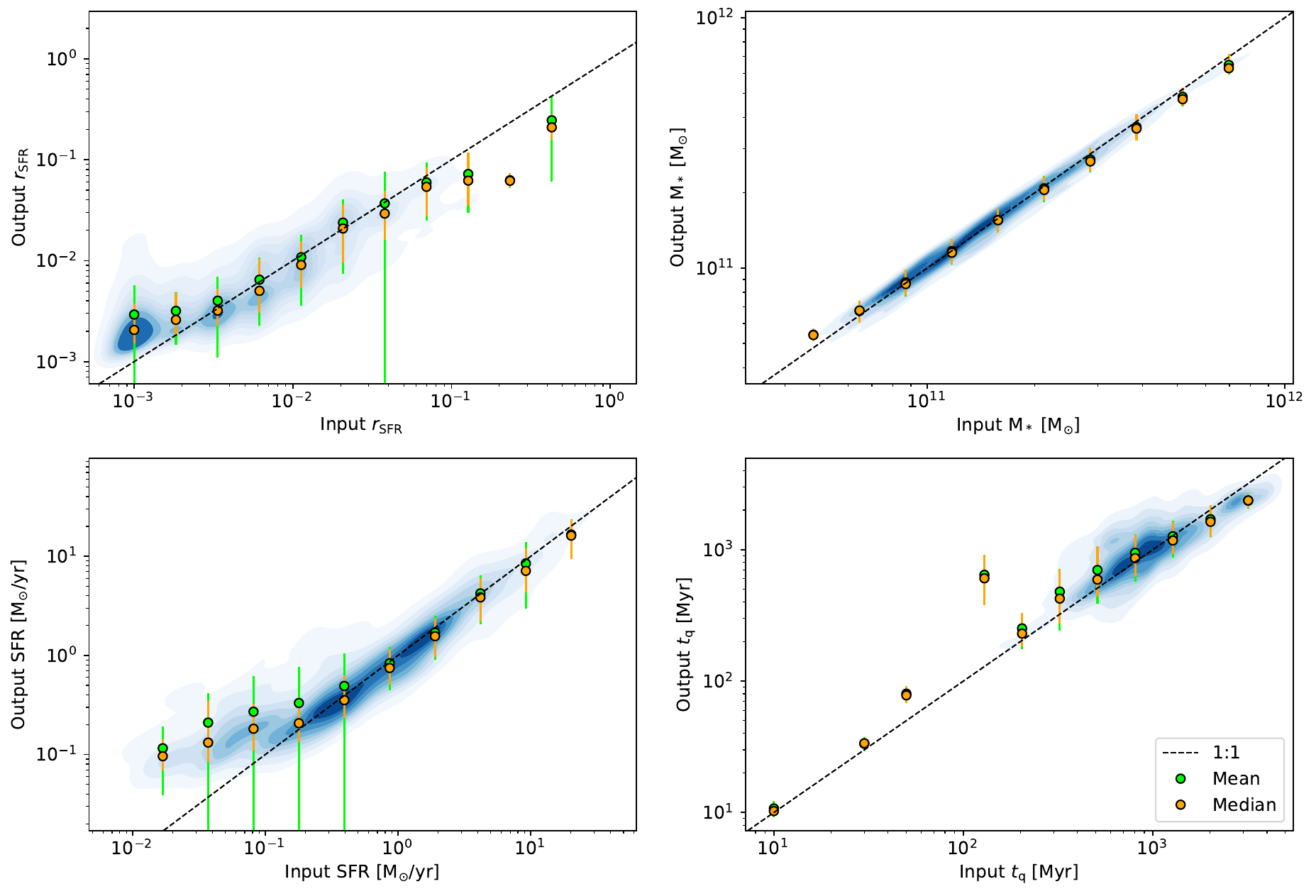}

\caption{Results of our \texttt{CIGALE} mock analysis: Bayesian output parameters in relation to the known input parameters, using 100 SEDs for each galaxy that were randomly disturbed in flux. The blue shaded areas show the density of individual values in the plot. Green points and error bars show the mean and standard deviation in bins of input parameters. Orange points and error bars show the median and the 16th and 84th percentiles in the same bins. The dashed line represents a one-to-one ratio.}
          \label{fig_cigale_mock_results}
\end{figure*}

Our \texttt{CIGALE} mock analysis follows the procedure described in \citet[][see also \citealt{boquien19cigale}]{ciesla21}.
First, we took the fluxes that \texttt{CIGALE} yields for the best-fit SED of each galaxy in its respective input filters. 
Each flux was then disturbed randomly 100 times, with a $\sigma$ corresponding to the original observational uncertainty, thus creating 100 mock observations of each galaxy with known parameters.
We then ran \texttt{CIGALE} again on this mock sample with the same parameter ranges as used previously.
In this analysis, we found a very good agreement between the Bayesian output parameters from \texttt{CIGALE} and the known input parameters, as shown in Fig.~\ref{fig_cigale_mock_results}. 
SFRs are very well constrained at intermediate and high input SFRs and only slightly overestimated (between +0.25 and +0.7\,dex) at SFRs $\lesssim0.2\,M_{\odot}$/yr; stellar masses, $t_{\mathrm{q}}$, and $r_{\mathrm{SFR}}$ are overall very well constrained across the entire dynamic range probed, with typical offsets of -0.01\,dex, +0.05\,dex, and -0.1\,dex, respectively, and dispersions well within the output uncertainties. 
Overall, this mock analysis supports the reliability of the galaxy parameters derived using \texttt{CIGALE}.

\FloatBarrier

\newpage
\section{Weight outliers} \label{app_weight_outliers}

\begin{figure}[h]
\centering
\includegraphics[width=0.6\linewidth]{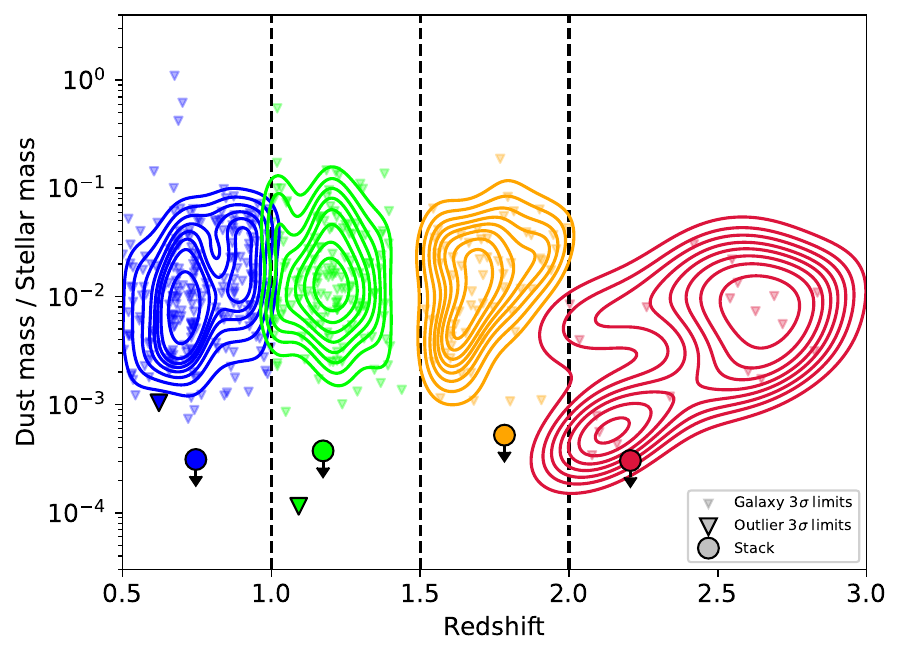}

\caption{Dust mass fraction inferred for all individual QGs in our sample for different redshift bins.
All QGs are individually undetected; therefore, triangles show the inferred $3\,\sigma$ upper limits.
The two high-weight galaxies are indicated by the large solid triangles and have redshifts of $z\sim0.6$ and $z\sim1.1$, respectively.
Circles show the upper limits of the stacks in each bin as shown in Fig.~\ref{fig_fdust_multiplot_onlyall} (i.e. without the high-weight sources).
The coloured contours represent the number density distribution within each redshift bin.
}
          \label{fig_indiv_measurements}
\end{figure}

In our sample of QGs, two galaxies stand out with significantly higher stacking weights (i.e. better observed sensitivity). 
Although these galaxies are not individually detected (like all other galaxies in our sample) they would strongly dominate the stacked signals of the respective redshift bins that contain them.
In Fig.~\ref{fig_indiv_measurements} we show the dust mass fraction for all QGs in our sample, which was inferred by performing aperture photometry on the ALMA images. 
The upper limits derived from stacking are generally lower than those inferred from individual galaxies, as is expected due to the improved depth of the stacks.
The two high-weight galaxies fall in the $z=0.5-1$ and $z=1-1.5$ bins.
In the $z=0.5-1$ bin, the high-weight galaxy falls on the lower edge of the $f_{\mathrm{dust}}$ distribution. 
Including this galaxy would have skewed the stack towards it, but the upper limit on $f_{\mathrm{dust}}$ inferred from the stack would not have been significantly different.
However, in the $z=1-1.5$ bin, the single high-weight galaxy yields an individual dust mass limit that is a factor of $\sim$3 lower than that of the stack, and a factor of $\sim$10 lower than that of the next lowest individual galaxy. 
This one yields such a low dust mass fraction since it is undetected in a deep ALMA project \citep[ASPECS-LP;][]{gonzales-lopez19,gonzales-lopez20} and very massive ($\log(M_*/M_{\odot})\sim11.7$). 
Including this galaxy in the stack would have yielded an $f_{\mathrm{dust}}$ upper limit almost identical to that of this individual galaxy.
While this QG is very constraining and therefore individually interesting, it is also not representative of the sample and, hence, needs to be excluded. 

\FloatBarrier

\section{Influence of 24\,$\mu$m-detected QGs} \label{app_analysis_24}

\begin{figure}[ht]
\centering
\includegraphics[width=0.6\linewidth]{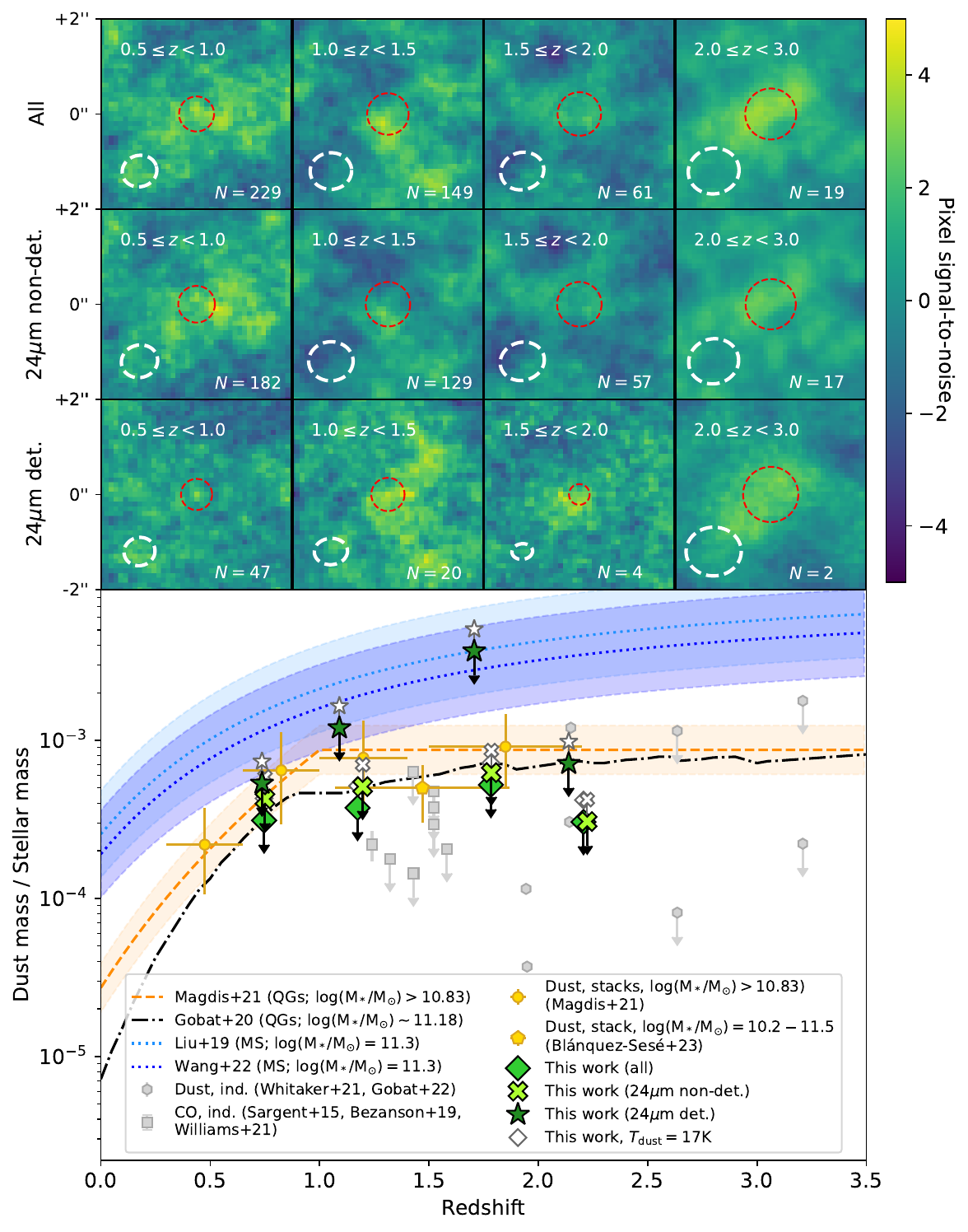}

\caption{Same as Fig.~\ref{fig_fdust_multiplot_onlyall} but now also showing stacking results for split samples, dividing galaxies with individual detections (dark green stars) from those with no detections (light green crosses) at 24\,$\mu$m.}
          \label{fig_fdust_lines_24}
\end{figure}

To investigate whether QGs that were individually detected in the \textit{Spitzer}/MIPS 24\,$\mu$m band have significantly different dust masses compared to those with undetected 24\,$\mu$m fluxes, we split the stacks accordingly (Fig.~\ref{fig_fdust_lines_24}). 
We find no significant difference. 
Because the MIPS-detected galaxies constitute only $\sim$16\% of the total sample, their stack upper limits on the dust mass fraction are statistically higher than those of the remaining sample.
Their limits for the intermediate-redshift bins extend into the MS regime, but it remains unclear whether these MIPS-bright galaxies truly differ from the average QG population, as these measurements remain upper limits.

\begin{figure}
\centering
\includegraphics[width=0.6\linewidth]{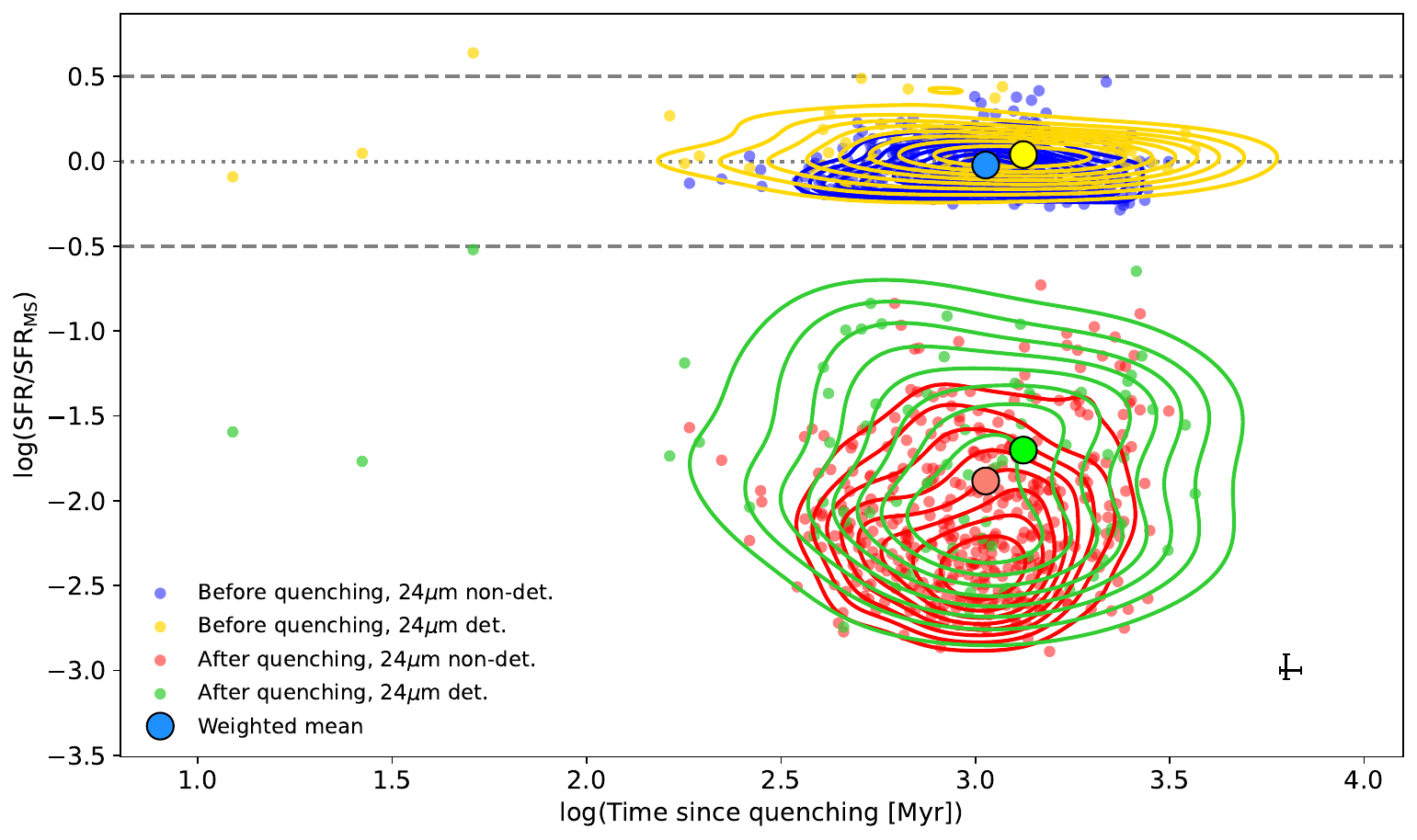}

\caption{Location with respect to the \citet{schreiber15} MS as a function of the time since quenching for our full QG sample. Green (red) dots mark the locations of galaxies detected (undetected) in the 24\,$\mu$m band after quenching, and yellow (blue) dots show the location of the same galaxies before quenching. The large symbols show the weighted mean of each group. The cross in the bottom-right corner denotes the error on the weighted mean. The coloured lines show the density distribution of each group.}
          \label{fig_deltaMS_agebq_24}
\end{figure}

We further investigated the groups of 24\,$\mu$m-detected and undetected QGs using the galaxy parameters yielded by \texttt{CIGALE} (Sect.~\ref{subsec_cigale_fitting}).
Figure~\ref{fig_deltaMS_agebq_24} shows $\Delta\mathrm{MS}_{\mathrm{before\ quenching}}$ and $\Delta\mathrm{MS}_{\mathrm{obs}}$ as a function of the time since quenching for 24\,$\mu$m-detected and -undetected QGs in our sample. 
Before quenching, we find no difference between the two groups with respect to position within the MS.
At the time of observation, the distribution of 24\,$\mu$m-detected QGs extends to somewhat higher values of $\Delta$MS than for 24\,$\mu$m-undetected QGs, but still remains well below the MS. 
We find no significant difference between the weighted means of the two groups.
The mean quenching times agree within 0.1\,dex, close to 1\,Gyr. 
The mean SFR of the 24\,$\mu$m-detected QGs is $\sim50\%$ (i.e. $\sim0.2$\,dex) higher than that of the 24\,$\mu$m-undetected QGs. 
This tentatively hints that these QGs might be quenched less efficiently compared to the rest of the population, retaining potentially a slightly larger gas and dust reservoir. 
There is, however, no indication that these objects could be mis-identified SFGs contaminating our sample.

\FloatBarrier

\section{Rest-frame wavelengths} \label{app_restframe_wavelength}

\begin{figure*}[h]
\centering
\includegraphics[width=0.88\linewidth]{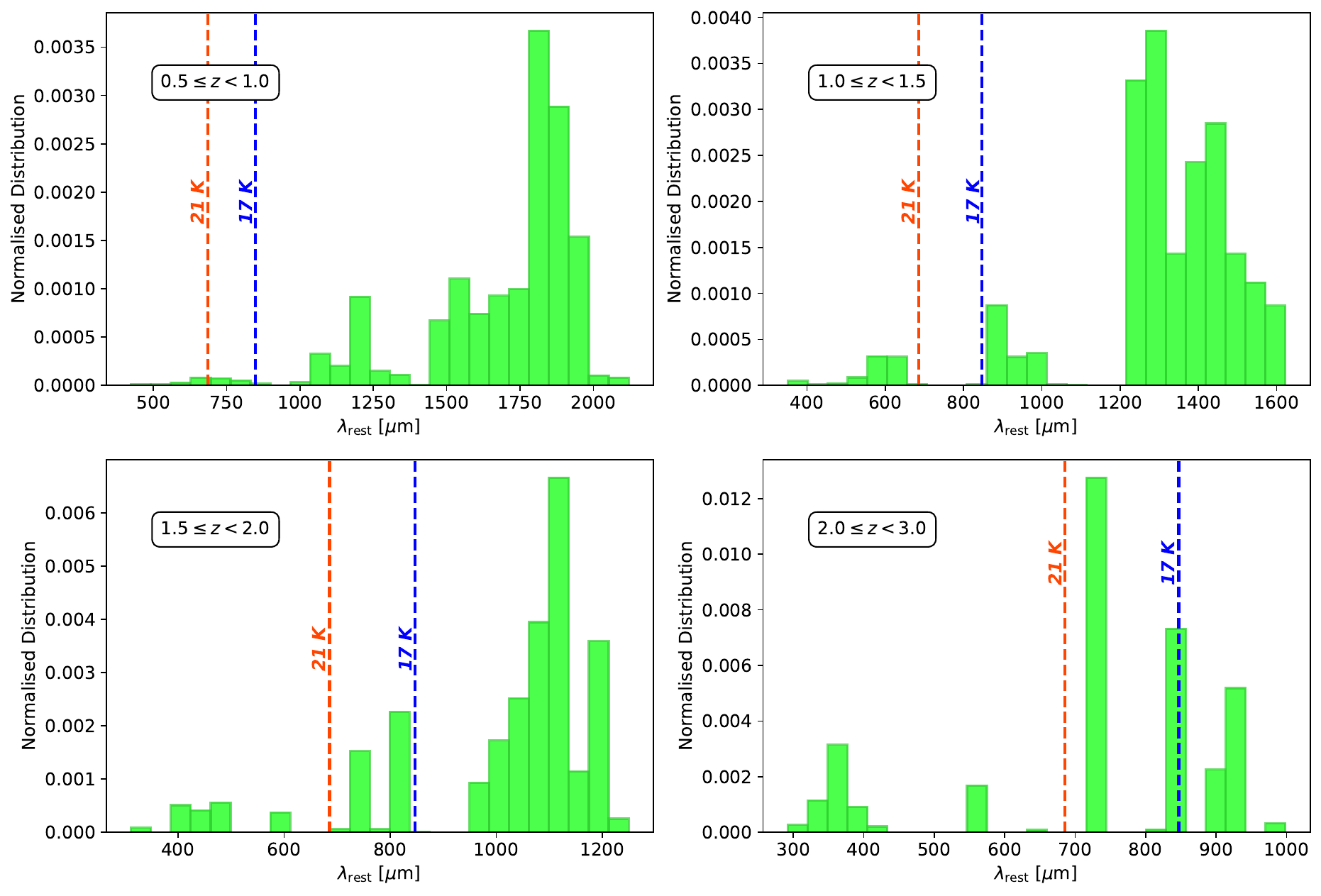}

\caption{Normalised distribution of rest-frame wavelengths used to compute $L_{850}^{\mathrm{rest}}$ for each stack in Fig.~\ref{fig_fdust_multiplot_onlyall}, weighted by the relative contribution of the individual pointing to the stacks.
The vertical dashed orange and blue lines indicate $\lambda=\frac{hc}{k_{\mathrm{B}}T}$ for a dust temperature of 21\,K and 17\,K, respectively.}
          \label{fig_restframe_wavel}
\end{figure*}

In Fig.~\ref{fig_restframe_wavel} we show the distribution of the rest-frame wavelengths in the stacks of Fig.~\ref{fig_fdust_multiplot_onlyall}, with vertical lines indicating the lower wavelength limit of the Rayleigh-Jeans spectral range, that is, $\lambda= \frac{hc}{k_{\rm B}T}$, for dust temperatures of 21\,K and 17\,K.
Our stacks are clearly dominated by the Rayleigh-Jeans regime in redshift bins from 0.5 to 2, which contain the highest number of galaxies. 
Therefore, our wavelength rescaling of luminosities (see Sect.~\ref{subsec_stacking}) is equivalent to a power-law rescaling \citep[with an assumed dust emissivity index of $\beta=1.8$;][]{magdis21} and thus largely independent of the exact dust temperature, even when assuming low, yet reasonable temperatures of 17\,K \citep{cochrane22}.
In the redshift range $>2$, the dominant part of the rest-frame wavelengths lie still within the Rayleigh-Jeans limit at $21$\,K, but move out for lower dust temperatures. 
Therefore, this particular bin is more sensitive to the chosen SED template and dust temperature.
Naturally, a significant deviation of $\beta$, $\Delta\beta$, from the assumed value of 1.8 could introduce an error into our rescaling.
Accounting for this effect, we found that for our four redshift bins it would translate into uncertainties that would range from $\sim1$\% to $\sim7$\% for $\Delta\beta=0.1$ \citep[i.e. a conservative dust emissivity dispersion measurement; e.g.][]{planck11}, and from $\sim10$\% to $\sim50$\% for $\Delta\beta=0.6$ \citep[i.e. a pessimistic dust emissivity dispersion measurement; e.g.][]{bendo25}.

\FloatBarrier
\newpage
\section{Acknowledgements}

We would like to thank the anonymous referee for their comments that helped to improve the paper.
SA and FB gratefully acknowledge the Collaborative Research Center 1601 (SFB 1601 sub-project C2) funded by the Deutsche Forschungsgemeinschaft (DFG, German Research Foundation) – 500700252.
LC acknowledges support from the French government under the France 2030 investment plan, as part of the Initiative d’Excellence d’Aix-Marseille Université – A*MIDEX AMX-22-RE-AB-101.
This paper makes use of the following ALMA data:
ADS/JAO.ALMA\#2015.1.00055.S, ADS/JAO.ALMA\#2015.1.00098.S, ADS/JAO.ALMA\#2015.1.00137.S, ADS/JAO.ALMA\#2015.1.00207.S, ADS/JAO.ALMA\#2015.1.00228.S, ADS/JAO.ALMA\#2015.1.00260.S, ADS/JAO.ALMA\#2015.1.00379.S, ADS/JAO.ALMA\#2015.1.00543.S, ADS/JAO.ALMA\#2015.1.00853.S, ADS/JAO.ALMA\#2015.1.00861.S, ADS/JAO.ALMA\#2015.1.00870.S, ADS/JAO.ALMA\#2015.1.00948.S, ADS/JAO.ALMA\#2015.1.01074.S, ADS/JAO.ALMA\#2015.1.01205.S, ADS/JAO.ALMA\#2015.1.01379.S, ADS/JAO.ALMA\#2015.1.01447.S, ADS/JAO.ALMA\#2015.1.01590.S, ADS/JAO.ALMA\#2016.1.00171.S, ADS/JAO.ALMA\#2016.1.00279.S, ADS/JAO.ALMA\#2016.1.00324.L, ADS/JAO.ALMA\#2016.1.00463.S, ADS/JAO.ALMA\#2016.1.00564.S, ADS/JAO.ALMA\#2016.1.00646.S, ADS/JAO.ALMA\#2016.1.00790.S, ADS/JAO.ALMA\#2016.1.00798.S, ADS/JAO.ALMA\#2016.1.00804.S, ADS/JAO.ALMA\#2016.1.00932.S, ADS/JAO.ALMA\#2016.1.00967.S, ADS/JAO.ALMA\#2016.1.00990.S, ADS/JAO.ALMA\#2016.1.01001.S, ADS/JAO.ALMA\#2016.1.01012.S, ADS/JAO.ALMA\#2016.1.01040.S, ADS/JAO.ALMA\#2016.1.01079.S, ADS/JAO.ALMA\#2016.1.01155.S, ADS/JAO.ALMA\#2016.1.01208.S, ADS/JAO.ALMA\#2016.1.01355.S, ADS/JAO.ALMA\#2016.1.01426.S, ADS/JAO.ALMA\#2016.1.01454.S, ADS/JAO.ALMA\#2016.1.01546.S, ADS/JAO.ALMA\#2016.1.01604.S, ADS/JAO.ALMA\#2017.1.00046.S, ADS/JAO.ALMA\#2017.1.00138.S, ADS/JAO.ALMA\#2017.1.00300.S, ADS/JAO.ALMA\#2017.1.00326.S, ADS/JAO.ALMA\#2017.1.00373.S, ADS/JAO.ALMA\#2017.1.00428.L, ADS/JAO.ALMA\#2017.1.00755.S, ADS/JAO.ALMA\#2017.1.00893.S, ADS/JAO.ALMA\#2017.1.01027.S, ADS/JAO.ALMA\#2017.1.01099.S, ADS/JAO.ALMA\#2017.1.01163.S, ADS/JAO.ALMA\#2017.1.01176.S, ADS/JAO.ALMA\#2017.1.01217.S, ADS/JAO.ALMA\#2017.1.01359.S, ADS/JAO.ALMA\#2017.1.01512.S, ADS/JAO.ALMA\#2017.1.01618.S, ADS/JAO.ALMA\#2017.1.01677.S, ADS/JAO.ALMA\#2017.1.01713.S, ADS/JAO.ALMA\#2017.A.00032.S, ADS/JAO.ALMA\#2017.A.00034.S, ADS/JAO.ALMA\#2018.1.00164.S, ADS/JAO.ALMA\#2018.1.00231.S, ADS/JAO.ALMA\#2018.1.00251.S, ADS/JAO.ALMA\#2018.1.00329.S, ADS/JAO.ALMA\#2018.1.00478.S, ADS/JAO.ALMA\#2018.1.00543.S, ADS/JAO.ALMA\#2018.1.00681.S, ADS/JAO.ALMA\#2018.1.00874.S, ADS/JAO.ALMA\#2018.1.00938.S, ADS/JAO.ALMA\#2018.1.01044.S, ADS/JAO.ALMA\#2018.1.01079.S, ADS/JAO.ALMA\#2018.1.01128.S, ADS/JAO.ALMA\#2018.1.01136.S, ADS/JAO.ALMA\#2018.1.01225.S, ADS/JAO.ALMA\#2018.1.01594.S, ADS/JAO.ALMA\#2018.1.01739.S, ADS/JAO.ALMA\#2018.1.01824.S, ADS/JAO.ALMA\#2018.1.01841.S, ADS/JAO.ALMA\#2018.1.01852.S, ADS/JAO.ALMA\#2018.1.01871.S, ADS/JAO.ALMA\#2019.1.00102.S, ADS/JAO.ALMA\#2019.1.00459.S, ADS/JAO.ALMA\#2019.1.00477.S, ADS/JAO.ALMA\#2019.1.00652.S, ADS/JAO.ALMA\#2019.1.00678.S, ADS/JAO.ALMA\#2019.1.00702.S, ADS/JAO.ALMA\#2019.1.00909.S, ADS/JAO.ALMA\#2019.1.01127.S, ADS/JAO.ALMA\#2019.1.01142.S, ADS/JAO.ALMA\#2019.1.01201.S, ADS/JAO.ALMA\#2019.1.01286.S, ADS/JAO.ALMA\#2019.1.01528.S, ADS/JAO.ALMA\#2019.1.01537.S, ADS/JAO.ALMA\#2019.1.01600.S, ADS/JAO.ALMA\#2019.1.01615.S, ADS/JAO.ALMA\#2019.1.01634.L, ADS/JAO.ALMA\#2019.1.01722.S, ADS/JAO.ALMA\#2019.2.00118.S. 
ALMA is a partnership of ESO (representing its member states), NSF (USA) and NINS (Japan), together with NRC (Canada), NSTC and ASIAA (Taiwan), and KASI (Republic of Korea), in cooperation with the Republic of Chile. The Joint ALMA Observatory is operated by ESO, AUI/NRAO and NAOJ.

\end{appendix}

\end{document}